\documentclass[prb,aps,reprint]{revtex4-1}

\usepackage{amsmath,amssymb,amstext,color}
\usepackage{feynmf}
\usepackage{txfonts}

\usepackage{lmodern}

\usepackage{graphicx}

\usepackage[hang, nooneline]{subfigure}

\usepackage{epstopdf}

\definecolor{DarkRed}{rgb}{0.65,0,0}
\definecolor{DarkGreen}{rgb}{0,0.55,0}
\definecolor{DarkBlue}{rgb}{0,0,0.9}
\definecolor{DarkOrange}{RGB}{238,113,25}

\definecolor{BrightRed}{rgb}{1,0.7,0.7}
\definecolor{BrightGreen}{rgb}{0.6,0.9,0.6}
\definecolor{BrightBlue}{rgb}{0.7,0.7,1}
\definecolor{BrightOrange}{RGB}{255,183,95}

\newcommand\varpm{\mathbin{\vcenter{\hbox{%
  \oalign{\hfil$\scriptstyle+$\hfil\cr
          \noalign{\kern-.3ex}
          $\scriptscriptstyle({-})$\cr}%
}}}}

\newcommand{\vect}[1]{\boldsymbol{#1}}
  
\newcommand{\im}{\mathrm i}
  
\newcommand{\e}[1]{\, \mathrm e^{\, #1 }}

\newcommand{\cone}[1]{\mathit{#1}}

\newcommand{\tauhat}[1]{\hat{\tau}_{#1}}

\newcommand{\diffint}[1]{\!\mathrm{d}#1 \,}

\begin{document}

%
%

\title{Spin Hall Effect and Spin Swapping in Diffusive Superconductors}
\author{Camilla Espedal$^{1}$, Peter Lange$^1$, Severin Sadjina, A.~G. Mal'shukov$^2$, and 
Arne Brataas$^1$}
\affiliation{$^1$ Department of Physics, Norwegian University of Science and Technology,
NO-7491 Trondheim, Norway\\
$^2$ Institute of Spectroscopy, Russian Academy of
Sciences, 142190, Troitsk, Moscow oblast, Russia}

\begin{abstract}
  We consider the spin-orbit-induced spin Hall effect and spin swapping in diffusive superconductors.   By employing the non-equilibrium Keldysh Green's function technique in the quasiclassical approximation, we derive coupled transport equations for the spectral spin and particle distributions and for the energy density in the elastic scattering regime.  We compute four contributions to the spin Hall conductivity, namely, skew scattering, side-jump, anomalous velocity, and the Yafet contribution. The reduced density of states in the superconductor causes a renormalization of the spin Hall angle. We demonstrate that all four of these contributions to the spin Hall conductivity are renormalized in the same way in the superconducting state. In its simplest manifestation, spin swapping transforms a primary spin current into a secondary spin current with swapped current and polarization directions. We find that the spin-swapping coefficient is not explicitly but only implicitly affected by superconducting correlations through the renormalized diffusion coefficients. We discuss experimental consequences for measurements of the (inverse) spin Hall effect and spin swapping in four-terminal geometries. In our geometry, below the superconducting transition temperature, the spin-swapping signal is increased an order of magnitude while changes in the (inverse) spin Hall signal are moderate. 
\end{abstract}

\maketitle

%
%

  \section{Introduction}
  \label{sec:introduction}
    
    The coupling between a quasiparticle's spin and its momentum causes an initially unpolarized current in conductors to become spin dependent. The resulting spin-orbit-induced effects can be intrinsic or extrinsic. Intrinsic effects are due to the manifestation of spin-orbit coupling in the quasiparticle band structure in combination with spin-conserving scattering events. Extrinsic effects are due to spin-orbit scattering off impurities. We focus on extrinsic effects that give rise to spin relaxation, spin swapping, spin Hall and inverse spin Hall effects.
    
    The simplest manifestation of the spin-orbit interaction is spin relaxation\cite{Yafet1963,Zutic2004}. This causes a nonequilibrium spin polarization to decay with time or an injected spin current to decay with distance. Below the superconducting transition temperature, measurements of the temperature dependence of the spin relaxation length can be used to determine the ratio between spin-orbit-induced and magnetic-impurity-induced spin relaxation.\cite{Poli2008} Our focus is on how the spin Hall effect and the spin swapping are affected by superconducting correlations.
    
    The correlation between the momentum and spin directions in the impurity scattering process can cause an injected primary spin current to transform into a secondary transverse spin current, even in the absence of electric (charge) currents. This effect is called spin swapping. In its simplest manifestation, the secondary current flows along the polarization of the injected current and with a polarization direction that is along the primary current flow - the spin currents have been `swapped'. This effect was first studied theoretically for extrinsic spin-orbit coupling.\cite{Lifshits2009} More recently, an intrinsic (Rash\-ba spin-orbit-induced) spin swapping effect\cite{Sadjina2012} was identified in two-dimensional diffusive metals. Spin swapping driven by electric fields in these systems has also been considered\cite{Shen2015}. We will determine how the spin swapping differs in the superconducting state compared to the normal state.
    
    The spin Hall effect has attracted considerable attention \cite{Karplus1954,Dyakonov1971b,Dyakonov1971,Hirsch1999,Zhang2000,Murakami2003,Sinova2004,Kato2004,Sih2005,Wunderlich2005,Saitoh2006,Engel2007,Stern2006,Valenzuela2006,Kimura2007,Jungwirth2012,Sinova2015}. There are two main contributions to the extrinsic spin Hall effect: skew scattering due to the spin-dependent quasiparticle scattering cross-section and the side-jump mechanism that arises from a spin-dependent displacement during the scattering events. Calculating the side-jump contribution to the spin Hall effect is a subtle issue because several terms contribute to this contribution.  In the stationary regime and in the absence of a magnetic field, we study three contributions in detail.
    
    The onset of superconductivity can renormalize the various spin transport effects and introduce new phenomena. The temperature dependence of the spin transport parameters below the critical temperature of the superconductor can be used to identify and quantify the competing spin-orbit-induced effects\cite{Poli2008}. A giant enhancement of the spin signal of up to five orders of magnitude in the superconducting state was reported experimentally\cite{Poli2008} in a nonlocal measurement setup. In niobium, there are measurements of a factor of four enhancement of the spin relaxation time in the superconducting state compared to the normal state\cite{Wakamura2014}. 
    
    In the inelastic transport regime, a giant increase in the nonlocal spin and charge accumulation signal due to the spin Hall effect was computed\cite{Takahashi2002, Takahashi2012} at low temperatures. Moreover, these studies indicated that the magnitudes of the skew scattering and the side-jump contributions are renormalized by different amounts below the superconducting critical temperature in spin Hall devices.  Recent non-local measurements found an inverse spin Hall signal that is 2000 times stronger in the superconducting state compared to the normal state\cite{Wakamura2015}.
    
    Quasiparticle transport is elastic when the quasiparticle energy is conserved during the scattering events. In the opposite regime, quasiparticle interactions cause transport to be inelastic, and the nonequilibrium distribution of the quasiparticles approaches equilibrium Fermi distributions that may be position, spin, and energy dependent. Spin transport in normal metals typically does not differ in the inelastic and elastic transport regimes since the temperature is considerably smaller than the relevant energy scale, that is, the Fermi energy. However, in superconductors, the typical temperatures are on a considerably smaller energy scale, namely that of the superconducting gap, and (spin) transport in the elastic and inelastic transport regimes can significantly differ\cite{Belzig2000,Tserkovnyak2002}. Since inelastic scattering rates increase with temperature, it is plausible that transport below the superconducting critical temperature is elastic\cite{Tserkovnyak2002}.  
    
    In this paper, we study the elastic transport of spin, particle, and energy in a diffusive superconductor. For this purpose, we use Keldysh nonequilibrium Green's functions. We include scattering from impurities, taking the spin-orbit coupling into account. We also complement our results with known effects of magnetic impurity scattering. We compute the renormalization of the spin Hall effect and spin swapping effect below the superconducting critical temperature. In contrast to recent theoretical works\cite{Takahashi2002, Takahashi2012} on inelastic scattering effects on transport in superconductors, we find the same renormalizations of all spin Hall contributions in the elastic transport regime. Moreover, we extend these studies to arbitrary spin polarizations and provide a rigorous discussion on the various contributions to the side-jump mechanism, including the anomalous velocity, the Yafet term, and an additional expression in the self-energy. Thus far, there have been no studies on the spin-swapping effect in superconductors. We demonstrate that the spin-swapping coefficient is only implicitly renormalized by superconducting correlations via the renormalized diffusion coefficients.
    
    We apply our transport formalism to study the (inverse) spin Hall and spin-swapping effects in a four-terminal geometry. In this geometry, we demonstrate that the signal resulting from the spin swapping can become an order of magnitude larger in the superconducting state compared to the normal state. On the other hand, change in the signal resulting from the (inverse) spin Hall effect are only moderate.
    
    The remainder of this paper is organized as follows. In Sec.~\ref{sec:transport}, we first present the microscopic Hamiltonian and the resulting transport equations for spin, particle, and energy transport, including scattering from magnetic and nonmagnetic impurities and from spin-orbit coupling. Sec.~\ref{sec:nonlocal} presents the four-terminal geometry and the calculation of the signals that result from the spin Hall effect and the spin swapping mechanism. Subsequently, Sec.~\ref{sec:derivation} provides an overview of the microscopic derivation of our results and a discussion of the side-jump mechanism and its contributions to the spin Hall effect. Finally, we present our conclusions in Sec.~\ref{sec:conclusion}. The appendices contain more details of our calculations.
    

%
%
    
  \section{Transport Equations}  
  \label{sec:transport}
  
    Let us first describe the microscopic model of the superconductor and then our primary results. The main results are the relation between the currents and the quasi-particle distributions and the diffusion equations. 
    
We describe the system using a four-component basis vector in spin $\otimes$ particle-hole space. We use a 'hat' to label vectors and matrices in this $4 \times 4$ space. The basis vector is

\begin{equation}
\label{eq:basis}
	\hat{\psi}^\dagger = (\psi_\uparrow^\dagger, \psi_\downarrow^\dagger,  \psi_\uparrow,  \psi_\downarrow) \, ,
\end{equation}    
    where $\psi_\sigma$ is the field annihilation operator for spin $\sigma$. 
    
    The field operators are described by the BCS one-particle Hamiltonian
    
    \begin{equation}
    \label{eq:hamiltonian}
    \hat{\mathcal{H}}(\cone{1}) =
      - \frac{\hbar^2}{2m} D_i (\cone{1})D_i (\cone{1}) + e\phi(\cone{1}) -\mu + \hat{\Delta}(\cone{1}) + \hat{U}_\text{tot} \, ,
    \end{equation}
    where $\phi$ is the scalar potential, $\mu$ is the chemical potential, $\hat{\Delta}(\cone{1})$ describes superconducting correlations, and $\hat{U}_\text{tot}$ describes both spin-conserving and spin-orbit-induced impurity scattering. We use an abbreviated notation for the coordinates that includes both spatial and temporal coordinates, where $\cone{1} = (\vect{r}_{\cone{1}},t_{\cone{1}})$.
    
   The kinetic energy in Eq.\ (\ref{eq:hamiltonian}) is expressed in terms of the co-variant derivative, 
\begin{equation}
\label{eq:covariant_derivative}
		D_\mu (\cone{1}) \hat{f}(\cone{1}) = \frac{\partial \hat{f}(\cone{1})}{\partial \cone{1}^\mu} + \im\tauhat{3} A_\mu (\cone{1}) \hat{f}(\cone{1})\, ,
\end{equation}	    
    where $\tauhat{3} = \text{diag} (1,1,-1,-1)$ is a generalization of the third Pauli matrix and $A_\mu \equiv (\phi/c,-\vect{A}) e/\hbar$ contains the scalar potential $\phi$ and the electromagnetic vector potential $\vect{A}$. 
	We also introduce its conjugate operator
\begin{equation}
\label{eq:covariant_derivative_conjugate}
		\hat{f}(\cone{1})D^\dagger_\mu (\cone{1})  = \frac{\partial \hat{f}(\cone{1})}{\partial \cone{1}^\mu} - \im \hat{f}(\cone{1}) \tauhat{3} A_\mu (\cone{1}) \, .
\end{equation}	       
 We use the standard four-vector notation, where Greek letters refers to both spatial and temporal coordinates, whereas Latin letters only take values referring to the spatial coordinates, i.e.  $\mu = 0, 1, 2, 3$ and $i=1,2,3$. Summation over repeated indices is implied.

	 Superconducting correlations are included via
    \begin{equation}
    \label{eq:gap_matrix}
      \hat{\Delta}(\cone{1})
      =
      \begin{pmatrix}
       0 & 0 & 0 & \Delta(\cone{1}) \\
       0 & 0 & -\Delta(\cone{1}) & 0 \\
       0 & \Delta^*(\cone{1}) & 0 & 0 \\
       -\Delta^*(\cone{1}) & 0 & 0 & 0 \\
      \end{pmatrix}
    \end{equation}
    which contains the s-wave superconducting scalar order parameter $\Delta(\cone{1}) = \lambda(\vect{r}_\cone{1}) \langle \psi_\downarrow(\cone{1}) \psi_\uparrow(\cone{1}) \rangle$, where
    $\lambda$ is the interaction strength and $\langle \dots \rangle$ denotes a quantum statistical average. 

The local potential
    \begin{equation}
    \label{eq:potentials_U}
      \hat{U}_\text{tot}(\vect{r})
      =
      \hat{U}(\vect{r})
            +
      \hat{U}_\text{so}(\vect{r})
    \end{equation}
    includes elastic impurity scattering and spin-orbit coupling. We express elastic impurity scattering as
    \label{eq:potentials}
    \begin{equation}
    \label{eq:potentials_imp}
      \hat{U}(\vect{r})
      =
      \sum_i
      u(\vect{r} - \vect{r}_i)
      ,
    \end{equation}
    where $u(\vect{r} - \vect{r}_i)$ is the $i^{\text{th}}$ elastic scattering potential at position $\vect{r}_i$.     We consider extrinsic spin-orbit  scattering governed by the impurities. The extrinsic spin-orbit coupling is described by
    \begin{align}
    \label{eq:potentials_soc}
    \hat{U}_\text{so}(\vect{r}) & =
      \sum_i \hat{u}_\text{so}(\vect{r} - \vect{r}_j) \\
      & =  \im\gamma \sum_j \big( \tauhat{3} \hat{\vect{\alpha}} \times \vect{\nabla} u(\vect{r} - \vect{r}_j) \big)^i  D_i(\vect{r})\nonumber
      ,
    \end{align}
%
    where $\hat{\vect{\alpha}} = \text{diag} (\bar{\vect{\sigma}},\bar{\vect{\sigma}}^{*})$ is a generalized vector of $4 \times 4$ Pauli matrices in electron-hole space. We denote the vector of conventional $2\times2$ Pauli matrices by $\bar{\vect{\sigma}}$. $\gamma$ is the spin-orbit interaction strength\footnote[1]{Here, we choose the sign such that spin-orbit coupling corresponds to its vacuum expression for which $\gamma_\text{vac} > 0$.}.

    The spin-orbit coupling at impurities can be understood as a consequence of the effects of intrinsic spin-orbit coupling in the band structure. The latter renormalizes the interaction strength $\gamma$ from its vacuum value. The renormalization can be interpreted as a shift in the physical position operator,
    \begin{equation}
    \label{eq:position_operator}
      \vect{r}
      \rightarrow
      \hat{\vect{r}}_\text{eff}
      =
      \vect{r} + \hat{\vect{r}}_\text{so}
      ,
    \end{equation}
    such that, to the first order in $\gamma$,
    \begin{equation*}
      \hat{U}(\hat{\vect{r}}_\text{eff})
      =
      \hat{U}(\vect{r}) + \hat{U}_\text{so}(\vect{r})
      ,
    \end{equation*}
    where the spin- and velocity-dependent correction to the position operator
    \begin{equation}
    \label{eq:Yafet}
      \hat{\vect{r}}_\text{so}
      =
      -
      \gamma (\tauhat{3} \hat{\vect{\alpha}} \times \vect{p})
    \end{equation}
    is known as the Yafet term\cite{Yafet1963, Engel2007} or the anomalous coordinate, where $\vect{p}$ is the momentum. The Yafet term in Eq.\ (\ref{eq:Yafet}) also contributes to the spin Hall effect, as we will discuss in more detail below. 

With this in hand, we find that the equation of motion for the four-component basis vector $\hat{\psi}$ is   
    \begin{equation}
    \label{eq:fourvector_equation_of_motion}
		\left[\im\hbar  \tauhat{3} D_0(\cone{1}) - \hat{\mathcal{H}}(\cone{1}) \right]\hat{\psi}(\cone{1}) \,  ,
    \end{equation}
     and for its conjugate, $\hat{\psi}^\dagger$ 
    \begin{equation}
    \label{eq:fourvector_equation_of_motion_conjugate}
       \hat{\psi}^\dagger(\cone{1})\left[-
       \im \tauhat{3}  D^\dagger_0 (\cone{1})
       -
       \hat{\uppercase{\mathcal{H}}}'(\cone{1})
      \right]
      =
      0
      ,
    \end{equation}
       where the 'prime' means that the covariant derivatives should be replaced by its conjugated form. In other words, $ \hat{\uppercase{\mathcal{H}}}'$ is the same as $ \hat{\uppercase{\mathcal{H}}}$, except that we let $D_\mu \rightarrow D_\mu^\dagger$. 
  
  Starting from these equations of motion, we use the Keldysh Green's function formalism to obtain expressions that describe the quasiparticle transport of spin, particle, and energy. We employ the quasiclassical approximation, which is valid for length scales that are considerably larger than the Fermi wavelength, and then the diffusion approximation, which is applicable when the system is far greater than the mean free path. We now present and discuss the main results. A rigorous derivation for interested readers is included in Sec.~\ref{sec:derivation}.

%
%

%
%

  \subsection{Current Expressions}  
  \label{subsec:transport_currents}     
    
In the elastic transport regime, energy is conserved and transport can be described at each energy $\epsilon$ relative to the chemical potential in terms of spectral ({\it energy-dependent}) currents and distributions. Our main result consists of two parts:  i) the relations between the quasiparticle spectral currents and the spectral distributions and ii) the spectral diffusion equations. We discuss the spectral currents in this section and the spectral diffusion equations in the next section. 

The spectral currents are the spectral particle current $j_i(\epsilon)$, the spectral spin current $j_{ij}(\epsilon)$, the spectral energy current ${j_\epsilon}_i(\epsilon)$, and the spectral spin-energy current ${j_\epsilon}_{ij}(\epsilon)$. Spectral currents with one subindex are particle ($j_i$) or energy (${j_\epsilon}_i$) and flow along the $i$ direction. Quantities with two subindices describe spin ($j_{ij}$) or spin-energy (${j_\epsilon}_{ij}$) current, where the first index ($i$) is the direction of the current flow and the second ($j$) denotes the spin polarization direction. 

The corresponding spectral particle and energy distribution functions are $h(\epsilon)$ and $h^\epsilon(\epsilon)$, respectively. Similarly, the spectral  spin and spin-energy distribution functions are $h^s_j(\epsilon)$ and ${h}^{\epsilon s}_j(\epsilon)$, respectively, where the subindex (here, $j$) denotes the spin polarization direction.  

From the spectral densities and spectral currents introduced in this section, the relevant physical quantities can be extracted. For example, the electric current density is a sum of all the spectral particle currents, 
    \begin{subequations}
    \label{eq:observables}
    \begin{equation}   
    \label{eq:observables_current_T}
      j_i^\text{tot}(\vect{R})
      =
      -
      \frac{e N_0}{2} \int \diffint{\epsilon}{} j_i (\vect{R}, \epsilon)
      ,
    \end{equation}
    and the electroneutrality dictates that the electrostatic potential is\cite{Rammer1986}
    \begin{equation}   
    \label{eq:observables_density_T}
      e \phi(\vect{R})
      =
      - 
      \int \diffint{\epsilon}{}
      N_\text{S}(\vect{R}, \epsilon) h(\vect{R}, \epsilon)
      ,
    \end{equation}
    \end{subequations}
    where  $N_s$ is the renormalization of the density of states due to superconducting correlations, which are introduced and discussed further below. The expressions for the spin, energy, and spin-energy properties are similar.

To the first order in the spin-orbit coupling, we compute that there are three contributions to the spectral current, namely, the conventional diffusion and supercurrent terms $j^{(0)}$, the spin Hall effects $j^{(\text{sH})}$, and the spin-swapping effect $j^{(\text{sw})}$:
    \begin{equation}
    \label{eq:currents_general}
      j(\epsilon)
      =
      j^{(0)}(\epsilon) + j^{(\text{sH})}(\epsilon) + j^{(\text{sw})}(\epsilon)
      .
    \end{equation}
    We will now discuss these contributions to the spectral current. To the zeroth order in the spin-orbit interaction strength, the currents are well known (see, e.g. Ref.\ \onlinecite{morten2004spin},):
    \begin{subequations}
    \label{eq:currents_diffusion}
    \begin{align} 
    \label{eq:currents_diffusion_T}
      j_i^{(0)}
      &
      =
      -
      D_\text{p} \nabla_i h
      + j^{sc}_i h^{\epsilon}
      ,
      \\
    \label{eq:currents_diffusion_TS}
      j_{ij}^{(0)}
      &
      =
      -
      D_\epsilon \nabla_i h^s_j
      + j^{sc}_i h^{\epsilon s}_j 
      ,
      \\  
    \label{eq:currents_diffusion_L}
       {j_\epsilon}_i^{(0)}
      &
      =
      -
      \big[ D_\epsilon \nabla_i h^\epsilon 
      + j^{sc}_i (1 - h)\big]
      ,
      \\
    \label{eq:currents_diffusion_LS}
      {j_\epsilon}_{ij}^{(0)}
      &
      =
      -
       \big[ D_p \nabla_i h_j^{\epsilon s}
      + j_i^{sc} h_j^s \big]
      .
    \end{align}
    \end{subequations}
    The diffusion coefficients $D_\epsilon$ and $D_{\text{p}}$ are well known, and their energy dependencies are governed by the superconducting correlations\cite{morten2004spin} and the nonequilibrium state. One simple limit of these expressions is the normal state, where $D_\epsilon = D_\text{p} = D$, where $D$ is the diffusion constant. Another simple limit is the BCS approximation of a dirty superconductor with no pair-breaking processes: $D_\text{p} / D = \epsilon^2 / (\epsilon^2 - \Delta^2)$ and $D_\epsilon / D = 1$ for energies above the gap, $|\epsilon| > |\Delta|$. The current in Eq.\ (\ref{eq:currents_diffusion}) also includes a supercurrent $\vect{j}^{sc}$, which is proportional to the gradient of the superconducting phase. Microscopic expressions for the generalized diffusion constants $D_\text{p}(\epsilon)$ and $D_\epsilon(\epsilon)$ as well as the supercurrent $\vect{j}^{sc}$ in general out-of-equilibrium conditions are given in Sec.~\ref{subsec:currents_and_densities}.
    
    Let us now turn to the spin-orbit-induced corrections to the conventional spectral current, which is one of our new central results. To the first order in the spin-orbit coupling strength $\gamma$, we compute contributions that correspond to the spin Hall and the inverse spin Hall effects and to the spin-swapping effect. We find that the contributions to the spectral current due to the spin Hall and the inverse spin Hall effects are\footnotemark[2] 
    \begin{subequations}
    \label{eq:currents_sh}
    \begin{align}
    \label{eq:currents_sh_T}
      j_i^\text{(sH)}
      &
      =
      -
      \chi_\text{sH}
      \varepsilon_{ijk}
      D \nabla_j (N_\text{S} h^s_k)
      ,
      \\
    \label{eq:currents_sh_TS}
      j_{ij}^\text{(sH)}
      &
      =
      \chi_\text{sH}
      \varepsilon_{ijk}
      D \nabla_k(N_\text{S} (h-1))
      ,
      \\
    \label{eq:currents_sh_L}
      {j_\epsilon}_i^\text{(sH)}
      &
      =
      -
      \chi_\text{sH}
      \varepsilon_{ijk}
      D \nabla_j (N_s h^{\epsilon s}_k)
      ,
      \\
    \label{eq:currents_sh_LS}
      {j_\epsilon}_{ij}^\text{(sH)}
      &
      =
      \chi_\text{sH}
      \varepsilon_{ijk}
      D \nabla_k (N_\text{S} h^\epsilon)
      ,
    \end{align}
    \end{subequations}
    where $N_\text{S}(\epsilon)$ is the ratio between the (energy-dependent) density of states in the superconducting state and the density of states in the normal state. The normal state spin Hall angle $\chi_\text{sH} = \chi_\text{sH}^\text{(sk)} + \chi_\text{sH}^\text{(sj)}$ is given in terms of the skew scattering constant,
    \begin{subequations}
    \label{eq:currents_constants_sh}
    \begin{equation}   
    \label{eq:currents_constants_rhosk}
      \chi_\text{sH}^\text{(sk)}
      =
      \frac{4 \eta}{3}
      \frac{\tau_\text{tr}}{\tau_\text{sk}}
      ,
    \end{equation}
    and the side-jump constant,
    \begin{equation}   
    \label{eq:currents_constants_rhosj}
      \chi_\text{sH}^\text{(sj)}
      =
      \frac{3 \gamma m}{\tau_\text{tr}}
      .
    \end{equation}
    \end{subequations}
    The dimensionless quantity $\eta = \gamma p_\text{F}^2 / 2$ is governed by the spin-orbit coupling strength, $p_\text{F}$ is the Fermi momentum, $\tau_\text{tr}$ is the transport relaxation time, and $\tau_\text{sk}$ is the skew scattering time.  

We find that the spin Hall angles that arise from skew scattering and side jump are all renormalized by equal amounts below the superconducting critical temperature via the renormalized density of states parametrized by $N_\text{S} (\epsilon)$.  In contrast, Ref. \onlinecite{Takahashi2012} computes the spin Hall conductivity in a different transport regime, the inelastic transport regime, and predicts that the renormalization of the spin Hall angle due to side jump and skew scattering differs. 

We note that both the spin Hall and inverse spin Hall effects described by Eq.\ (\ref{eq:currents_sh})  are created by quasiparticles, while contributions from the condensate are absent. The origin of this is that the inverse spin Hall effect is induced by a nonequilibrium spin accumulation governed by the distribution function, whereas the phase of the condensate  wave-function remains intact. This is in contrast to the equilibrium magnetoelectric effect produced by a static Zeeman field  in a spin-orbit-coupled superconductor discussed in Ref.\ \onlinecite{Zeeman}. In that case, the condensate current emerges due to mixing of spin-singlet and spin-triplet Cooper pairs. Such a situation could occur out of equilibrium by taking into account that an effective Zeeman field may be created by spin-polarized electrons due to the Coulomb exchange interaction of itinerant electrons \cite{Exchange}. Here, we assume that the Coulomb interaction is weak and disregard this effect. 

We also disregard the condensate supercurrent associated with the  conversion of quasiparticle current  into the supercurrent due to the inelastic relaxation of quasiparticles (the so-called charge imbalance relaxation). At low temperatures, such a relaxation occurs at large length 
scales. We assume that our system is small enough to disregard charge imbalance relaxation.

    When the scattering potential is isotropic, the transport relaxation time $\tau_\text{tr}$ equals the elastic scattering time $\tau$, $\chi_\text{sH}^\text{(sk)} = 4 \pi \eta N_0 u_0 / 3$, and $\chi_\text{sH}^\text{(sj)} = 3 \gamma m / \tau$, where $u_0 = u(\vect{q}=0)$ is the Fourier transformed scattering potential at $\vect{q} = 0$ and $N_0$ is the density of states at the Fermi level in the normal state. Our results in Eq.\ (\ref{eq:currents_sh}) are valid for general anisotropic scattering potentials, except that the skew scattering contribution~\eqref{eq:currents_constants_rhosk} is computed to the lowest order in small anisotropies; see Sec.~\ref{sec:derivation}. Several factors contribute to the side jump~\eqref{eq:currents_constants_rhosj}, and we discuss these factors in more detail in Sec.~\ref{sec:derivation} and Appendix~\ref{sec:sidejump}. 
    
    Let us study the superconductivity-induced renormalization of the spin Hall angle in the elastic transport regime in more detail. For this purpose, we consider a weakly perturbed superconductor in which the density of states is constant and equal to the BCS density of states. We can then express the spin Hall contribution to the spectral particle current, the inverse spectral spin Hall current, from the zeroth-order spin current,
    \begin{equation}
        {j}_i^\text{(sH)} =\theta_{\text{sH}}\varepsilon_{ijk} {j}_{jk}^\text{(0)} \, , 
    \end{equation}
    where we have defined the spin Hall angle in the superconducting state as $\theta_{\text{sH}} = \chi_\text{sH} N_{\text{s}}(D/D_\epsilon)$. Whereas $(D_\epsilon/D)$ typically only weakly depends on the energy, $N_{\text{s}}$ strongly varies as a function of energy. In the superconducting state, $N_\text{S}$ is greatly enhanced close to the superconducting gap, which causes a significant increase of the spin Hall angle $\theta_{\text{sH}}$ . The fact that there is a giant enhancement in the spin Hall angle for quasiparticles with energies around the gap is consistent with the main findings of Refs. \onlinecite{Takahashi2012} and \onlinecite{Wakamura2015}. We provide microscopic expressions for the density of states in the superconductor with respect to its normal state value $N_s(\epsilon)$ and the scattering times in Eq.~\eqref{eq:currents_constants_sh} in Sec.~\ref{sec:derivation}.
    
    The spin-swapping effect\cite{Lifshits2009, Sadjina2012} couples only spins. To display the spin swapping current in a compact manner, we define the operator $[\vect{a},\vect{b}]^{(sw)}_{ij} \equiv \delta_{ij}a_c b_c - a_j b_i$, and we obtain the spectral currents
    \begin{subequations}
    \label{eq:currents_sw}
    \begin{align}
    \label{eq:currents_sw_TS}
      j_{ij}^\text{(sw)}
      &
      =
      \chi_{\text{sw}} D_\epsilon[\vect{\nabla} ,\vect{h}^s]^{(sw)}_{ij}
      +\frac{\chi_{\text{sw}}}{2}[(\vect{\nabla} D_\epsilon) ,\vect{h}^s]^{(sw)}_{ij}
      \\
      &+ \frac{\chi_{sH}}{2} 
      [N_s\vect{j}^{sc,2} + \vect{\mathcal{R}}^p, \vect{h}^{\epsilon s}]^{(sw)}_{ij}
      \nonumber ,
      \\
    \label{eq:currents_sw_LS}
     {j_\epsilon}_{ij}^\text{(sw)}
      &
      =
      \chi_{\text{sw}} D_p[\vect{\nabla} ,\vect{h}^{\epsilon s}]^{(sw)}_{ij}
      +\frac{\chi_{\text{sw}}}{2}[(\vect{\nabla} D_p) ,\vect{h}^{\epsilon s}]^{(sw)}_{ij}
      \\
      &+ \frac{\chi_{sH}}{2} 
      [N_s\vect{j}^{sc,2} + \vect{\mathcal{R}}^p, \vect{h}^{\epsilon s}]^{(sw)}_{ij}
      \nonumber
      .
    \end{align}
    \end{subequations}
    The normal state spin-swapping constant is\cite{Lifshits2009}
    \begin{equation}   
    \label{eq:currents_constants_chi}
      \chi_\text{sw}
      =
      \frac{4 \eta}{3}
      \frac{\tau_\text{tr}}{\tau_\text{sw}}
      ,
    \end{equation}
    where $\tau_\text{sw}$ is the spin-swapping scattering time. The spin-swapping constant reduces to $\chi_\text{sw} = 4 \eta / 3$ when the scattering potential is isotropic. In its simplest manifestation, spin swapping interchanges the direction of flow and the spin polarization direction, as follows from Eq.~\eqref{eq:currents_sw}. A prominent feature of the spin-swapping effect is that it leads to in-plane spin polarizations at the lateral edges of a two-dimensional sample, whereas the spin Hall effect gives rise to out-of-plane spin polarizations, which makes it possible to experimentally distinguish the two effects. We find that spin swapping is renormalized in the superconducting state only through the generalized diffusion constants contained in the diffusion currents in Eq.~\eqref{eq:currents_sw}, while $ \chi_\text{sw}$ remains unchanged by superconducting correlations. 
    
    The additional terms in Eq.~\eqref{eq:currents_sw} appear when there are spatial variations in the magnitude and phase of the superconducting order parameter. The term proportional to $\vect{j}^{sc,2}$ can be viewed as super-spin-swapping current. In addition to this term, we have a more complicated term that is related to the gradient of $\theta$, which is related to gradients in the spectral properties of the superconductor.
    
    The expressions for the spectral currents, Eqs.~\eqref{eq:currents_sh} and~\eqref{eq:currents_sw}, satisfy Onsager's reciprocal relations. For example, the spin Hall effect and the inverse spin Hall effect are governed by the same susceptibility $\chi_{\text{sH}}$.

The spin Hall effect and the spin swapping mechanisms can be detected in nonlocal geometries. In these setups, the detected signals will also depend on the counterflow of  currents due to spin and particle distribution build-ups. We will subsequently compute these effects and the resulting effect of the superconducting correlations on the electrochemical potentials that can be detected.


%
%

  \subsection{Diffusion Equations}  
  \label{subsec:transport_diffusion}    
    
   We now turn to the presentation of the spectral ({\it energy-dependent}) diffusion equation. 
    We find that the diffusion of particle, spin, and energy is described in terms of energy-dependent diffusion equations:\footnote[2]{We use the Einstein notation throughout this paper, i.e.\ repeated indices are implicitly summed over.}
    \begin{subequations}
    \label{eq:diffusion}
    \begin{align}   
    \label{eq:diffusion_T}
      \nabla_i j_i
      &
      =
      \alpha h + \alpha^\epsilon h^{\epsilon} 
      ,
      \\
    \label{eq:diffusion_TS}
      \nabla_i j_{ij}
      &
      =
      \Big(
	\frac{\alpha_\text{so}}{\tau_\text{so}}
	+
	\frac{\alpha_\text{m}}{\tau_\text{m}}
      \Big)
      h_j
      ,
      \\
    \label{eq:diffusion_L}
      \nabla_i {j_\epsilon}_i
      &
      =
      0
      ,
      \\
    \label{eq:diffusion_LS}
      \nabla_i {j_\epsilon}_{ij}
      &
      =
      \alpha^\epsilon h_j^s
      +
      \alpha h^{\epsilon s}_j
      +
      \Big(
	\frac{{\alpha_\epsilon}_\text{so}}{\tau_\text{so}}
	+
	\frac{{\alpha_\epsilon}_\text{m}}{\tau_\text{m}}
      \Big)
      {h}^\epsilon_j
      ,
    \end{align}
    \end{subequations}

    The terms proportional to $\alpha$ and $\alpha^\epsilon$ in Eqs.~\eqref{eq:diffusion_T} and~\eqref{eq:diffusion_LS} are proportional to the superconducting gap and are responsible for converting quasiparticle currents into supercurrents.\cite{Schmid1975} For completeness, we have also included the effects of the magnetic impurities, where we use the results from Ref.\ \onlinecite{morten2004spin}. The spin relaxation terms in Eqs.~\eqref{eq:diffusion_TS} and~\eqref{eq:diffusion_LS} are given in terms of the spin relaxation scattering times $\tau_\text{so}$ and $\tau_\text{m}$ due to spin-orbit coupling and magnetic impurities, respectively. Eq.~\eqref{eq:diffusion_L} expresses that spectral energy is conserved in the elastic transport regime. 
    
    Superconducting correlations lead to the introduction of the renormalization factors $\alpha$, $\alpha_\text{so}$, $\alpha_\text{m}$, ${\alpha_\epsilon}_\text{so}$, and ${\alpha_\epsilon}_\text{m}$ \cite{morten2004spin}. These factors are energy dependent and are governed by the superconducting state. Microscopic expressions for these renormalization factors (and the scattering times) are presented in Sec.~\ref{sec:derivation}. To obtain insights into how the various effects occurring in Eq.~\eqref{eq:diffusion} are renormalized, let us consider a scenario in which the superconductor has properties that are close to that of a bulk BCS superconductor (BCS limit). This is, for instance, realized in large superconductors that are weakly coupled to reservoirs that inject spin and particle currents. Quasiparticles can propagate for energies above the gap, $|\epsilon| > |\Delta|$, when there is no conversion of quasiparticle currents to supercurrents and $\alpha = 0$. At the same time, the spin relaxation renormalization factors become ${\alpha_\epsilon}_\text{so} = {\alpha_\epsilon}_\text{m} = \epsilon^2 / (\epsilon^2 - \Delta^2)$ for spin energy, and $\alpha_\text{so} = 1$ and $\alpha_\text{m} = (\epsilon^2 + \Delta^2) / (\epsilon^2 - \Delta^2)$ for spin density. This implies that the spin-orbit-induced spin relaxation rate is identical in the superconducting and normal states whereas the spin-energy relaxation rate is enhanced in the superconducting state. In contrast, when magnetic impurities dominate the relaxations of spins, both the spin relaxation rate and the spin-energy relaxation rates are enhanced for quasiparticles with energies above and close to the superconducting gap.
    
%
%

\section{Spin Transport in Non-local Geometries}
\label{sec:nonlocal}

\begin{figure}[h!tb]
    \centering
      \subfigure[]
	{
	\includegraphics[width=7.8cm]{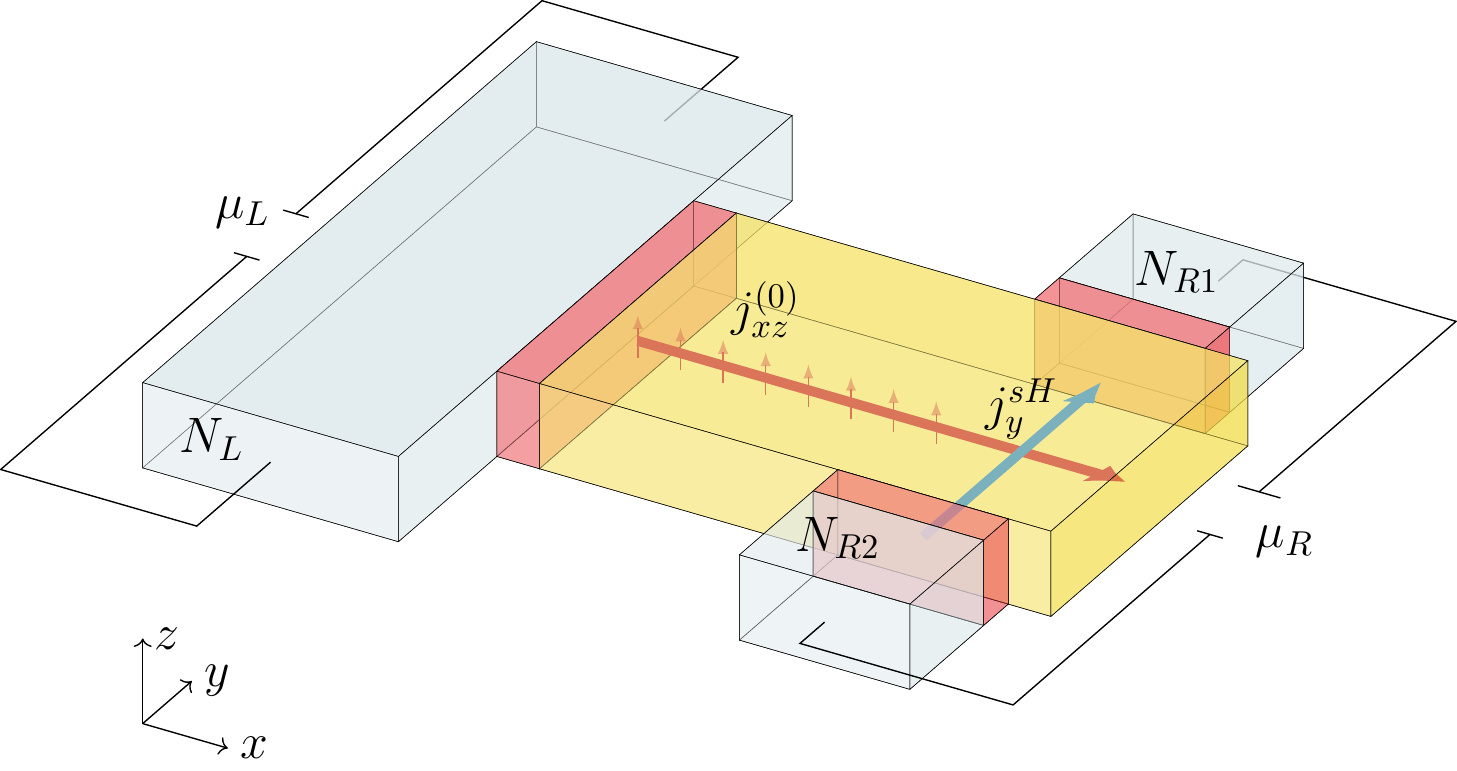}
	\label{fig:figure_ishe}
	}
      \subfigure[]
	{
	\includegraphics[width=7.8cm]{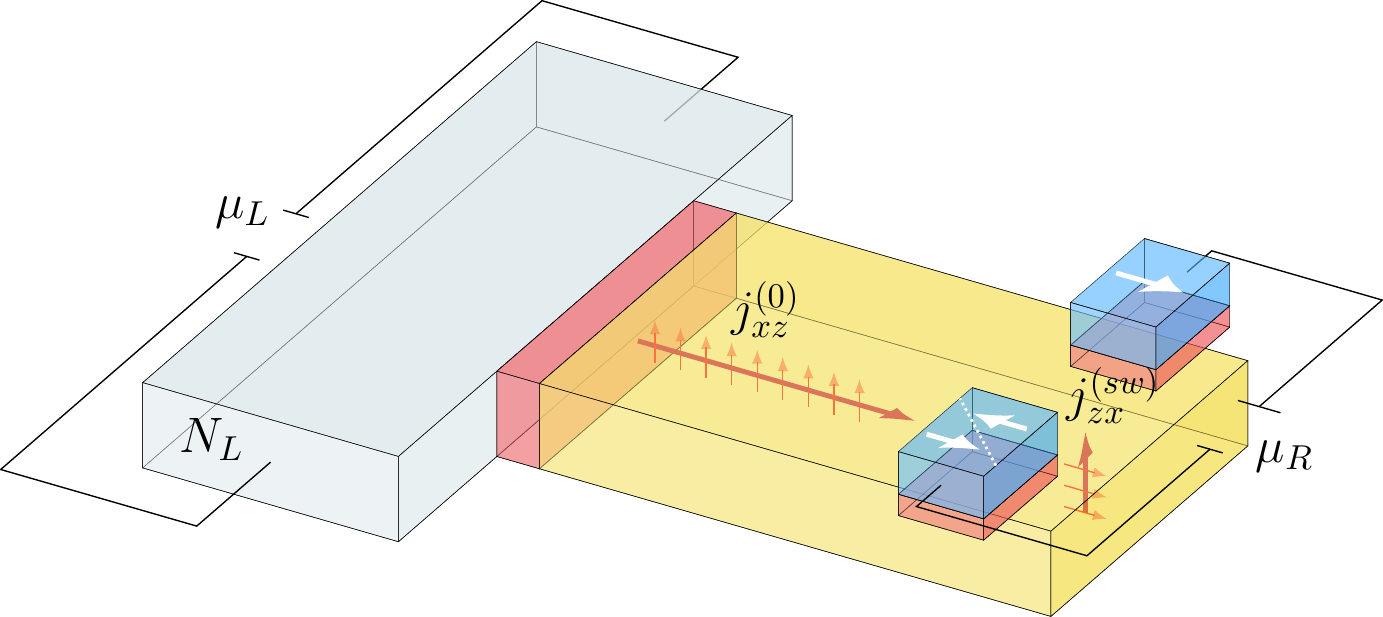}
	\label{fig:figure_swap}
	}
      \caption{
      	Non-local geometries for measuring the spin Hall effect (a) and the spin swapping (b). In both cases, a particle current flowing in the left normal metal generates via the spin Hall effect a spin current that flows into the superconductor. Inside the superconductor, the inverse spin Hall effect converts the spin current into a particle current, and spin swapping swaps the spin current polarization and flow directions. In (a), the electrochemical potentials in the normal metals measure the inverse spin Hall effect. In (b), the spin-polarized contacts can be switched between a parallel and anti-parallel configuration to detect the spin-swapping effect.
      }
    \label{fig:nonlocal_geometries}
    \end{figure}

We will now compute the signatures of the (inverse) spin Hall effect and the spin swapping in non-local geometries. We consider the setups in Fig.\ \ref{fig:nonlocal_geometries}. The left normal metal (N$_L$) functions as a spin injector into the superconductor (S) via a tunnel junction. The additional normal metals to the right (N$_{R1}$ and N$_{R2}$) act as detectors of the spin-particle-coupled properties of the superconductor. 

We assume that the transparency of the tunnel contacts is low such that there are no proximity effects between the normal metals and the superconductor. The equilibrium properties of the superconductor are then the same as if it were detached from the rest of the circuit. Furthermore, we assume that the resistances of the tunnel contacts used for detecting the inverse spin Hall and spin-swapping effects in the superconductors are considerably larger than the resistance of the tunnel contact used for spin injection. In this limit, we can first compute non-equilibrium the spin distribution in the normal metal, which is not influenced by the rest of the circuit. In turn, this spin distribution leaks into the superconductor. Finally, the electrochemical potential difference between the normal metals to the right (N$_{R1}$ and N$_{R2}$) detects the inverse spin Hall effect and spin swapping without influencing the spin and particle distributions in the superconductor.  Our geometry differs from the setup in \cite{Wakamura2014} since the spin current into the superconductor is injected along its long axis.

We first focus on the spin injection that originates from the left normal metal. Since the tunnel resistances are large, we can consider the properties of the left normal metal independently of the rest of the circuit. The normal metal is biased so in a way to in a way to maintain the particle distribution in the normal metal close to the tunnel contact in equilibrium. We consider that this is achieved with a electrochemical potential $-\mu_L/2$ at the top and $\mu_L/2$ at the bottom of the left normal metal. Such a setup prevents a charge imbalance from flowing into the superconductor. The total electric current through the normal metal is then $I=G \mu_L/e$, where the conductance is $G= e^2 N_L D_L A_L/L_L$ in terms of the density of states $N_L$, diffusion coefficient $D_L$, cross section $A_L$, and length $L_L$ of the left normal metal. The particle current flowing along the $y$-direction generates via the spin Hall effect a spin current  flowing along the $x$-direction that is spin polarized along the $z$-direction. In turn, the spin current induces a spin distribution at the edges of the left (L) normal metal. The standard calculation explained below shows that the spin distribution in the left normal metal close to the tunnel interface is
\begin{equation}
h_z^{s(L)}(\epsilon) = \zeta_{L} h_\text{eq}(\epsilon,\mu_L/2)\, . 
\label{eq:spinacc}
\end{equation}
where the dimensionless particle-spin conversion efficiency $\zeta_{L}$ is independent of the energy and 
\begin{equation}
h_\text{eq}(\epsilon,\mu_L/2)=\frac{1}{2} \left[ \tanh{\frac{\mu_L/2 + \epsilon}{2 k_B T}} + \tanh{\frac{\mu_L/2 - \epsilon}{2 k_B T}} \right]
\end{equation}
arises from the distributions of the quasi-particles in the reservoirs at electrochemical potentials $\mu_L/2$ and $-\mu_L/2$.

We compute the particle-spin conversion efficiency parameter $\zeta_L$ as follows. To the zeroth order in the spin-orbit coupling, we use Eq.\ (\ref{eq:currents_diffusion_T}) to find the relation between the spatial variation of the spectral particle distribution and the spectral particle current, $ j_y^{(0)}(\epsilon)=-D \partial_y h(\epsilon)$ and solve the diffusion equation of Eq. (\ref{eq:diffusion_T}) to find the spatially varying spectral particle distribution $h(\epsilon)$. To the first order in the spin-orbit coupling, the spatial variation of the spectral particle distribution gives rise to a spin current $j_{xz}(\epsilon)$ and an associated spin distribution $h_z^s(\epsilon)$ in the normal metal. The spatial variation of the spin distribution $h^s_z(\epsilon)$ is determined by the diffusion equation (\ref{eq:diffusion_TS}) with the boundary conditions that the spectral spin current vanishes at the edges of the normal metal. The spectral spin current is from Eq.\ (\ref{eq:currents_sh_TS}) $j_{xz}(\epsilon) = - D_L \partial_x h_z^s(\epsilon) - \chi_{\text{sH},L} D \partial_y h(\epsilon)$. Solving the diffusion equation (\ref{eq:diffusion_TS}) with these boundary conditions and assuming the normal metal is wider than its diffusion length, we find that the spin distribution of Eq.\ (\ref{eq:spinacc}) with with the particle-spin efficiency parameter $\zeta_L$ = $-2\chi_{\text{sH},L} \lambda_L/L_L$, which is a dimensionless property of the left normal metal spin injector where the spin-diffusion length is $\lambda_L$.

Next, we will compute the resulting spin and particle distribution in the superconductor. Since we will focus on spin-distribution-induced spin-particle conversion effects in the supercurrent, we only need to focus on how spins propagate from the left normal metal into the superconductor. At a low transmission interface, the spectral spin currents through the interfaces are
\begin{align}
j_{xz} = N_S(\epsilon) g_T (h_z^{s(N)}-h_z^s) \, , 
\end{align}
where the spin distribution at the normal metal side, $h_z^{s(N)}$, was computed in Eq.\ (\ref{eq:spinacc}). The conductance of the tunnel junction when the superconductor is in the normal state is $G_T=N_0 g_T$. We solve the spin-diffusion equation (\ref{eq:diffusion_TS}) in the superconductor with the boundary condition of continuity of the spin current. We also expand the result to the lowest order in the tunnel conductance and assume that the superconductor is considerably longer than the spin-diffusion length (along the $x$-direction). We then find that the spatially dependent spin distribution is governed by the ratio between the tunnel conductance $N_S(\epsilon) G_T$ and the conductance of the superconductor within the spin-flip length $l(\epsilon)$, $N_0 D_\epsilon(\epsilon)/\lambda(\epsilon)$:
\begin{equation}
h_z^s(x,\epsilon) =  \zeta_L h_\text{eq}(\epsilon,\mu_L/2)  \frac{g_T N_s(\epsilon) \lambda(\epsilon)}{D_\epsilon(\epsilon)}  e^{-x/\lambda(\epsilon)} \, .
\label{eq:spinaccS}
\end{equation}
The energy-dependent spin-flip length is  $\lambda(\epsilon) = [D_\epsilon(\epsilon) \tau_\text{sf}(\epsilon)]^{1/2}$, where the spin-flip relaxation rate is $1/\tau_\text{sf}=\alpha_\text{so}/\tau_\text{so} + \alpha_\text{m}/\tau_\text{m}$. 

In the following, we will show how the spatially dependent spin distribution of Eq.\ (\ref{eq:spinaccS}) in the superconductor gives rise to the inverse spin Hall effect and spin swapping.

\subsection{Inverse Spin Hall Effect}

In the inverse spin Hall geometry, the  inverse spin Hall effect is measured via normal metals  in tunnel contact with the superconductor. From Eq.\ (\ref{eq:currents_sh_T}), we see that the spin-Hall-induced spectral particle current density is 
\begin{equation}
j_y^\text{sH}(\epsilon) = \chi_{\text{sH},S} D N_S(\epsilon) \partial_x h_z^s(x,\epsilon) \, , 
\label{eq:specISH}
\end{equation}
where we computed $h_z^s(x,\epsilon)$ in Eq.\ (\ref{eq:spinaccS}). 

We assume that the transverse width $W_S$ of the superconductor is smaller than the charge-imbalance relaxation length. The spin-Hall-induced spectral particle current density of Eq.\ (\ref{eq:specISH}) must then be compensated by the zeroth-order spectral particle current density induced by a transverse gradient of the spectral particle distribution. Since transport is assumed to be elastic, we use the boundary conditions that the total (zeroth-order and spin Hall contributions) spectral particle current density should vanish at the lateral edges. We also only take into account the difference between the particle distributions at $y=W_S/2$ and $y=-W_S/2$ over which the potential is detected. The resulting relative spectral particle distribution at $y=W_S/2$  is then
\begin{equation}
h(\epsilon) = \eta_\text{psp} h_\text{eq}(\epsilon,\mu_L/2) \frac{D^2 N_s^2(\epsilon)}{2D_p(\epsilon) D_\epsilon(\epsilon)} e^{-x/\lambda(\epsilon)} \, ,
\label{eq:pspparticleaccumulation}
\end{equation}
where the dimensionless particle-spin-particle conversion is governed by $\eta_\text{psp}=- \zeta_L  \chi_{\text{sH},S} W_S g_T /D$. 

The particle distribution of Eq.\ (\ref{eq:pspparticleaccumulation}) can be detected as a electrochemical potential in another normal metal in tunnel contact with the superconductor. The spectral particle current between the superconductor and this tunnel contact is 
\begin{align}
j_{y} = N_S(\epsilon) g_{T'} \left[ h -h_\text{eq}(\epsilon,\mu_R/2) \right] \, , 
\end{align}
The electrochemical potential $\mu_R$ in the normal metal that we detect is determined by the integral equation that the total current into the right normal metal should vanish, $\int d\epsilon j_y(\epsilon)=0$, and is therefore indepedent of the detector tunnel conductance $g_{T'}$.

In linear response, we expand 
\begin{equation}
h_\text{eq}(\epsilon,\mu) \approx -[\partial_\epsilon f(\epsilon) - \partial_\epsilon f(-\epsilon)]\mu \, , 
\label{eq:hexpansion}
\end{equation}
 where $f(\epsilon)$ is the Fermi-Dirac distribution function. We then  find that the ratio between the electrochemical potentials in the superconducting state and the normal metal is
\begin{equation}
\label{eq:ishe_signal}
\frac{\mu^{(S)}}{\mu^{(N)}}=  \frac{ \int_\Delta^\infty d\epsilon~ e^{- x/\lambda(\epsilon)} \partial_\epsilon f(\epsilon) N_S(\epsilon)^3 [D / D_\epsilon(\epsilon)]}{e^{-x/\lambda_{\text{N}}}\int_\Delta^\infty d\epsilon \partial_\epsilon f(\epsilon) N_S(\epsilon)[D_p(\epsilon) D_\epsilon(\epsilon)/D^2]} \, ,
\end{equation}
where $\lambda_N$ is the spin-diffusion length in the normal state. The electrochemical potential when the superconductor is in its normal state is $\mu^{(N)}=\mu_R= \eta_\text{psp} e^{-x/\lambda_N}\mu_L$. Naturally, the electrochemical potential is proportional to the spin Hall angle in the left normal metal and the spin Hall angle in the superconductor via the particle-spin-particle conversion coefficient $\eta_\text{psp}$.  

We consider first the case when spin-flip is predominately due to spin-orbit scattering, $1/\tau_\text{so} \gg 1/\tau_\text{m}$. Remarkably, 
there is then an exact compensation of the factors in the numerator and denominator of Eq.\ (\ref{eq:ishe_signal}) so that $V^{(S)}=V^{(N)}$. This is because $N_S(\epsilon)^2=D_p(\epsilon)D_\epsilon(\epsilon)/D^2$ and $\lambda(\epsilon)=\lambda_N$ in that limit. This ensures that the particle imbalance of Eq. (\ref{eq:pspparticleaccumulation}) attains its normal state value even when superconducting correlations are taken into account.  

When spin-flip scattering due to magnetic impurities become stronger, there is a decay of the spin-Hall signal when the temperature is reduced below the superconducting transition temperature. This is caused by the reduction of the the spin-diffusion length $\lambda(\epsilon)$  with respect to its normal state value $\lambda_N$ in this regime.

We conclude that the inverse spin Hall signal is equal to or smaller than its value in the normal state below the superconducting transition temperature.

\subsection{Spin Swapping}

To study spin swapping, we consider the geometry shown in Fig.\ \ref{fig:nonlocal_geometries}b). In this scenario, spin swapping implies that the spin current flowing in the superconductor along the $x$-direction that is polarized along the $z$-direction will be swapped into a secondary (and smaller) spin current that flows along the $z$-direction and is polarized along the $x$-direction. From Eq.\ (\ref{eq:currents_sw_TS}), we  find that the spin-swapped-induced spectral spin current density is 
\begin{equation}
j_{zx}^{(sw)}(\epsilon) = -\chi_{\text{sw},S} D_\epsilon \partial_x h_z^s(x,\epsilon) \, . 
\label{eq:swapS}
\end{equation}
Maintaining that the transverse width of the superconductors is smaller than the charge imbalance length, the swapped spectral spin current of Eq.\ (\ref{eq:swapS}) must be counterbalanced by a spin current induced by a transverse gradient of the spin distribution. Requiring a vanishing spectral spin-current at the edges ($z=-d_S/2$ and $z=d_S/2$) then determined the transverse secondary swapped spin distribution 
\begin{equation}
h_x^{(sw)}(x,y,\epsilon) = \eta_\text{pss} \frac{\lambda(\epsilon)}{d_S} N_S(\epsilon) \frac{D}{D_\epsilon(\epsilon)} e^{-x/\lambda(\epsilon)}, ,
\label{eq:pssaccumulation}
\end{equation}
where $\eta_\text{pss}=\zeta_L g_T d_S \chi_{\text{sw},S}/D$ is a  dimensionsless particle-spin-spin conversion factor. 

We can already here note that the swapped spin distribution of Eq.\ (\ref{eq:pssaccumulation}) becomes larger for energies around the gap than in the normal state. As we will demonstrate below, this also leads to an enhanced spin swapping signal.

The detection of the spin swapping signal of Eq.\ (\ref{eq:pssaccumulation}) requires the use of spin-polarized contacts. Hence, we assume a setup such as the one shown in Fig.\ \ref{fig:nonlocal_geometries}, where the right tunnel contacts consist of ferromagnets with a spin polarization along the $x$-direction. We also assume that the magnetization of the tunnel contact can be made to be parallel or anti-parallel. Furthermore, to detect the spatial variation of the swapped spin distribution along the $z$-direction, we consider a situation in which the tunnel contacts are attached on top of the superconductor. In such an experiment, we can detect the swapped spin current. 

We detect the electrochemical potential in large probe reservoirs where there is no spin distribution. The particle distributions in the detectors attain their local equilibrium values $h_ {eq}(\epsilon,\mu_R)$ with respect to the detector electrochemical potential $\mu_R$. The spectral particle current through the detector spin-polarized tunnel barrier is
\begin{equation}
j_{z}^{\pm}(\epsilon) = N_S(\epsilon) g_{Td}  \left[\pm P_{Td}\left(-h_x^{(sw)} \right) + \left(h_\text{eq}(\epsilon,\mu^\pm)-h^{(S)} \right)  \right] \, , 
\end{equation}
where the sign $\pm$ indicates whether the tunnel polarizer is parallel or anti-parallel to the $x$-direction. $h^{(S)}$ is the particle distribution in the superconductor that will not play a role in our spin-swapping detection scheme. Requiring no total current to the reservoir, such that also $\int d\epsilon ( j_z^+ - j_z^-) = 0$, we find an expression for the electrochemical potential difference $\Delta \mu = \mu^+ - \mu^-$ in linear response by using the expansion of Eq.\ (\ref{eq:hexpansion}):
\begin{equation}
\Delta \mu =P_{Td} \eta_\text{css} \frac{ \int d\epsilon N_s(\epsilon) h_x^{sw}(\epsilon)}{\int d\epsilon N_s(\epsilon) \partial_\epsilon h_{Ld} (\epsilon)}  \, , 
\end{equation}
where we computed the transverse swapped spin distribution $h_x^{sw}$ in Eq.\ (\ref{eq:pssaccumulation}).
\begin{widetext}
It is instructive to consider the ratio between the electrochemical potential difference in the superconducting state versus the normal state
\begin{equation}
\label{eq:swap_signal}
	\frac{\Delta \mu^{(N)}}{\Delta \mu^{(S)}} =  \frac{\int_\Delta^\infty d\epsilon N_s(\epsilon)^2 \left[D/D_\epsilon(\epsilon)\right]\left[\lambda(\epsilon)/\lambda_N \right]\left[\partial_\epsilon f(\epsilon)\right] e^{-x/\lambda(\epsilon)}}{\int_\Delta^\infty d\epsilon N_s(\epsilon) \partial_\epsilon f(\epsilon) e^{-x/\lambda_N}}.
\end{equation}
\end{widetext}

\begin{figure}[h!tb]
\centering
	\includegraphics[width=7.8cm]{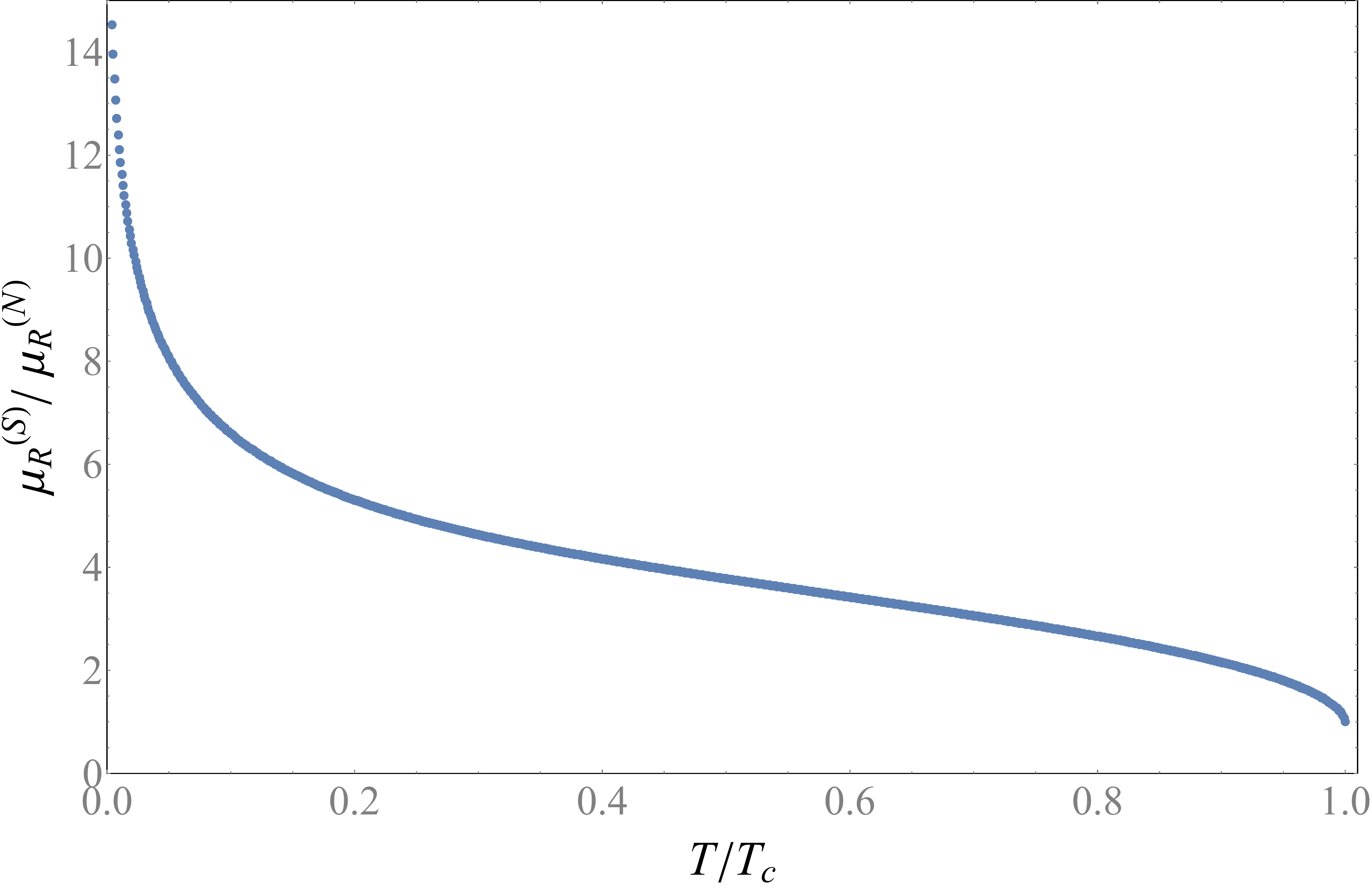}
	\caption{The temperature dependence of the spin-swapping signal.}
	\label{fig:swap_plot}
\end{figure}

We evaluate Eq.\ (\ref{eq:swap_signal}) numerically and the result is presented in Fig. \ref{fig:swap_plot}.
As announced, below the superconducting transition temperature, there is an enhancement of the spin-swapping signal. This can be understood from  Eq.\ (\ref{eq:pssaccumulation}) and Eq.\ (\ref{eq:swap_signal}). The spin-swapping spin distribution is enhanced for energies around the gap in the superconducting state. This leads to the enhancement of the spin-swapping signal at temperature below the superconducting transition temperature.

  \section{Microscopic Derivation}  
  \label{sec:derivation}
  	
  	\subsection{Definition of the Green's function}
  	\label{subsec:derivation_definition}

    In this section, we will derive our results presented in Sec.~\ref{sec:transport}, the diffusion equations (\ref{eq:diffusion}), and the relations between the currents and the distribution functions of Eqs.\ (\ref{eq:currents_diffusion}), (\ref{eq:currents_sh}), and (\ref{eq:currents_sw}). Our starting point is the microscopic Hamiltonian of Eq.~\eqref{eq:hamiltonian}, and we use the nonequilibrium Keldysh Green's function formalism.
    
    We define the kinetic Green's function in terms of the $4$-component vector of Eq.~\eqref{eq:basis}:
    \begin{subequations}
    \label{eq:greens_definition}
    \begin{equation}
    \label{eq:_greens_definition_K}
      \hat{G}_{ij}^\text{K}(\cone{1},\cone{1'})
      =
      - \im (\tauhat{3})_{ii} \big\langle \big[ \hat{\psi}_i (\cone{1\,}), \, \hat{\psi}^\dagger_j (\cone{1'}) \big]_{-} \big\rangle
      ,
    \end{equation}
    where $[A,B]_{\pm} = AB \pm BA$. $\hat{G}^\text{K}$ is a $4 \times 4$ matrix in spin~$\otimes$~particle-hole space, and we denote such matrices using a `hat' superscript. Similarly, we define the retarded Green's function,
    \begin{equation}
    \label{eq:greens_definition_R}
      \hat{G}_{ij}^\text{R}(\cone{1},\cone{1'})
      =
      - \im \Theta(t_\cone{1} - t_\cone{1'})
      (\tauhat{3})_{ii} \big\langle \big[ \hat{\psi}_i (\cone{1\,}), \, \hat{\psi}^\dagger_j (\cone{1'}) \big]_{+} \big\rangle
      ,
    \end{equation}
    and the advanced Green's function,
    \begin{equation}
    \label{eq:greens_definition_A}
      \hat{G}_{ij}^\text{A}(\cone{1},\cone{1'})
      =
      \im \Theta(t_\cone{1'} - t_\cone{1})
      (\tauhat{3})_{ii} \big\langle \big[ \hat{\psi}_i (\cone{1\,}), \, \hat{\psi}^\dagger_j (\cone{1'}) \big]_{+} \big\rangle
      ,
    \end{equation}
    \end{subequations}
    where $\Theta(t)$ is the Heaviside function. Next, we construct an $8 \times 8$ Green's function matrix $\check{G}(\cone{1},\cone{1'})$ in spin~$\otimes$~particle-hole~$\otimes$~Keldysh space,\cite{Keldysh1965}
    \begin{equation}
    \label{eq:greens_definition_matrix}
    \check{G} = 
    \begin{pmatrix}
      \hat{G}^\text{R} & \hat{G}^\text{K} \\
      \hat{0} & \hat{G}^\text{A}
      \end{pmatrix}
      ,
    \end{equation}
    which obeys the right-handed equation of motion 
    \begin{equation}
    \label{eq:greens_equation_of_motion_rh}
      \big(
	\im \hbar c \tauhat{3} D_0(\cone{1})
	-
	\hat{\uppercase{\mathcal{H}}}(\cone{1})
      \big)
      \check{G}(\cone{1},\cone{1'})
      =
      \delta(\cone{1} - \cone{1'})
    \end{equation}
    and its corresponding left-handed equation of motion
    \begin{equation}
    \label{eq:greens_equation_of_motion_lh}
    \check{G}(\cone{1},\cone{1'})
      \big(
	- \im \hbar c \tauhat{3} D'_0(\cone{1'})
	-
	\hat{\uppercase{\mathcal{H}}}' (\cone{1'})
      \big)
      =
      \delta(\cone{1} - \cone{1'})
    \end{equation}
    in terms of the Hamiltonian \eqref{eq:hamiltonian}. We denote $8 \times 8$ Green's function matrices using a `check' superscript. The operators to the left of the Green's function in Eq.~\eqref{eq:greens_equation_of_motion_rh} are $8 \times 8$ matrices decomposed into two identical block-diagonal $4 \times 4$ matrices that occupy the retarded and advanced components in Keldysh space.
    
	\subsection{Derivation of the co-variant Eilenberger equations}    
    \label{subsec:definitions_eilenberger}    

    The Eilenberger equation is widely used\cite{Rammer1986}. Nevertheless, we include a derivation of the Eilenberger equation for systems in which the extrinsic spin-orbit interaction is essential. Spin-orbit interactions require careful handling of the spin-orbit-induced self-energy that appears in the Eilenberger equation. The Eilenberger equation is obtained\cite{Rammer1986} by taking the difference between the left- and right-handed equations of motion, Eqs.~\ref{eq:greens_equation_of_motion_lh} and \ref{eq:greens_equation_of_motion_rh}. By taking the Wigner transform in the mixed representation and keeping terms to first the order in $\hbar$\cite{comment}, we obtain for a stationary system
	\begin{equation}
	\label{eq:wigner_transform_subtracted_dyson}
	\begin{split}
	&\im\hbar v_i \tilde{\nabla}_i \check{G} + [\epsilon \tauhat{3} + \hat{\Delta}, \check{G}]_- - [\hat{\Sigma}_{imp} - \hat{\Sigma}_{m}\stackrel{\otimes}{,}\check{G}]_- 
	\\ 
	&-(\hat{\Sigma}_{so}\otimes\check{G} - \check{G}\otimes\hat{\Sigma}'_{so}) = 0,
	\end{split}
	\end{equation}
	where $\tilde{\nabla}_\mu X \equiv \nabla_\mu X +~\im  A_\mu[\tauhat{3}, X]_-$ and $\hat{\Sigma}_{so}$ contains all the self-energy contributions involving $\hat{U}_{so}$. 
	
	    
        In a stationary state, the quasiclassical Green's function is defined as
    \begin{equation}
    \label{eq:greens_definition_quasiclassical}
      \check{g}(\vect{R},\vect{p}_\text{F}, \epsilon)
      =
      \frac{\im}{\pi}
      \int \diffint{\xi_{\vect{p}}}{}
      \check{G}_\text{c}(\vect{R}, \vect{p}, \epsilon)
      ,
    \end{equation}
    which in the mixed representation is a function of the center-of-mass coordinate $\vect{R} = (\vect{r}_\cone{1} + \vect{r}_\cone{1'})/2$ and the energy $\epsilon$ related to the relative time coordinate $t = t_\cone{1} - t_\cone{1'}$ by a co-variant Wigner transform as defined in Appendix~\ref{sec:fourier}. The momentum $\vect{p}_\text{F}$ is related to the relative position $\vect{r} = \vect{r}_\cone{1} - \vect{r}_\cone{1'}$ by a Fourier transform and is fixed at the Fermi level, and the integration variable is $\xi_{\vect{p}} = p^2 / 2m$. The quasiclassical Green's function \eqref{eq:greens_definition_quasiclassical} is determined by the Eilenberger equation which in a stationary state reads 
    \begin{equation}
    \label{eq:eilenberger}
    \begin{split}
      &0
      =
      \im \hbar \vect{v}_\text{F} \cdot \tilde{\nabla}\check{g}
      +
      \big[
	\tauhat{3} \epsilon + \hat{\Delta}, \check{g}
      \big]_{-} 
      - 
      \big[
	\check{\sigma}, \check{g}
      \big]_{-} 
      \end{split}
      ,
    \end{equation}
    where $\vect{v}_\text{F} = \vect{p}_\text{F}/m$ is the Fermi velocity, and we have inserted the various self-energies that we will address in the next section. Upon impurity averaging, $\hat{\Sigma}_{so}$ and $\hat{\Sigma}'_{so}$ become identical and are included in the commutator in  Eq.~\eqref{eq:eilenberger}. Eq.~\eqref{eq:eilenberger} does not determine the quasiclassical Green's function uniquely; therefore, we also need a normalization condition,\cite{Rammer2007}
    \begin{equation}
      \label{eq:normalization}
      \check{g}^2=1
      .
    \end{equation}

 We have now derived the Wigner-transformed Eilenberger equation in the presence of spin-orbit interactions.

	\subsection{Self-energies}
	\label{subsec:definitions_selfenergies}    
    
    We consider a diffusive system and will therefore compute average quantities relevant at length scales longer than the mean free path and independent of the impurity configuration. First, we include the effects of the local potential $\hat{U}_\text{tot}$ within the self-consistent Born approximation. To the second order in the local potential, the self-energy is shown in Fig.~\ref{fig:selfenergy_born_1} and reads as
    \begin{equation}
    \label{eq:selfenergy_definition_born}
      \check{\Sigma}(\cone{1}, \cone{1'})
      =
      \big\langle
	\hat{U}_\text{tot}(\vect{r}_\cone{1})
	\check{G}_\text{c}(\cone{1}, \cone{1'})
	\hat{U}_\text{tot}(\vect{r}_\cone{1'})
      \big\rangle_\text{c}
      ,
    \end{equation}
    where $\langle \dots \rangle_\text{c}$ denotes averaging over all possible impurity configurations, and we assume that $\langle \hat{U}_\text{tot} \rangle_\text{c} = 0$. The self-energy $\check{\Sigma}$ is a functional of the impurity-averaged full propagator $\check{G}_\text{c} = \langle \check{G} \rangle_\text{c}$. The terms in the local potential give rise to various self-energy contributions that can be treated independently. In the absence of spin-orbit coupling, the effects of elastic impurity scattering are calculated from
    \begin{subequations}
    \label{eq:selfenergy_definition}
    \begin{equation}
    \label{eq:selfenergy_definition_imp}
      \check{\Sigma}_\text{imp}(\cone{1}, \cone{1'})
      =
      n
      \int \diffint{\vect{r}_i}{}
      u(\vect{r}_\cone{1} - \vect{r}_i)
      \check{G}_\text{c}(\cone{1}, \cone{1'})
      u(\vect{r}_\cone{1'} - \vect{r}_i)
      ,
    \end{equation}
    where $n$ is the impurity concentration. 
    Spin swapping and side-jump scattering arise from contributions that are linear in the spin-orbit coupling strength:
    \begin{equation}
    \label{eq:selfenergy_definition_so1}
    \begin{split}
      \check{\Sigma}_\text{so}^{(1)}(\cone{1}, \cone{1'})
      &=
      n
      \int \diffint{\vect{r}_i}{}
      \hat{u}_\text{so}(\vect{r}_\cone{1} - \vect{r}_i)
      \check{G}_\text{c}(\cone{1}, \cone{1'})
      u(\vect{r}_\cone{1'} - \vect{r}_i)
      \\
      &+
      n
      \int \diffint{\vect{r}_i}{}
      u(\vect{r}_\cone{1} - \vect{r}_i)
      \check{G}_\text{c}(\cone{1}, \cone{1'})
      \hat{u}_\text{so}(\vect{r}_\cone{1'} - \vect{r}_i)
      .
    \end{split}
    \end{equation}
    \begin{figure}[h!tb]
    \centering
      \subfigure[]
	{
	\includegraphics[width=4cm]{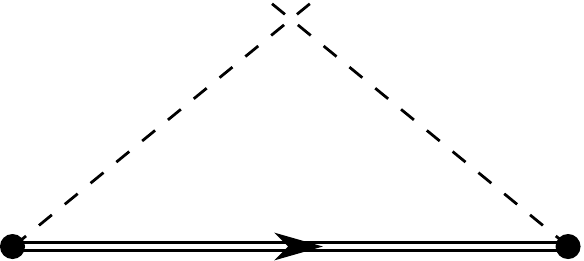}
	\label{fig:selfenergy_born_1}
	}
      \subfigure[]
	{
	\includegraphics[width=4cm]{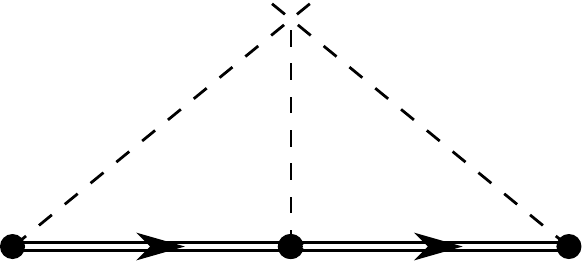}
	\label{fig:selfenergy_born_2}
	}
      \caption{Self-energy diagrams. \subref{fig:selfenergy_born_1}~Self-consistent Born approximation. \subref{fig:selfenergy_born_2}~Third-order contribution that determines skew scattering.}
    \label{fig:selfenergy_born}
    \end{figure}
    To include spin-orbit-induced spin relaxation, we also calculate the second-order contributions from the spin-orbit coupling to the self-energy:
    \begin{equation}
    \label{eq:selfenergy_definition_so}
      \check{\Sigma}_\text{so}(\cone{1}, \cone{1'})
      =
      n
      \int \diffint{\vect{r}_i}{}
      \hat{u}_\text{so}(\vect{r}_\cone{1} - \vect{r}_i)
      \check{G}_\text{c}(\cone{1}, \cone{1'})
      \hat{u}_\text{so}(\vect{r}_\cone{1'} - \vect{r}_i)
      .
    \end{equation}
    
    We will also include contributions from skew scattering to the spin Hall and inverse spin Hall effects. However, skew scattering does not appear within the framework of the self-consistent Born approximation. To include skew scattering, we also include contributions that are third order in the potential $u$: see Fig.~\ref{fig:selfenergy_born_2}.\cite{Kohn1957, Engel2007} To the first order in the spin-orbit coupling, the skew scattering contributions to the self-energy are
    \begin{widetext}
    \begin{equation}
    \label{eq:selfenergy_definition_skew}
    \begin{split}
      \check{\Sigma}_\text{sk}(\cone{1}, \cone{1'})
      &=
      n
      \int \diffint{\vect{r}_i}{}
      \int \diffint{\cone{2}}{}
      \hat{u}_\text{so}(\vect{r}_\cone{1} - \vect{r}_i)
      \check{G}_\text{c}(\cone{1}, \cone{2})
      u(\vect{r}_\cone{2} - \vect{r}_i)
      \check{G}_\text{c}(\cone{2}, \cone{1'})
      u(\vect{r}_\cone{1'} - \vect{r}_i)
      \\
      &+
      n
      \int \diffint{\vect{r}_i}{}
      \int \diffint{\cone{2}}{}
      u(\vect{r}_\cone{1} - \vect{r}_i)
      \check{G}_\text{c}(\cone{1}, \cone{2})
      \hat{u}_\text{so}(\vect{r}_\cone{2} - \vect{r}_i)
      \check{G}_\text{c}(\cone{2}, \cone{1'})
      u(\vect{r}_\cone{1'} - \vect{r}_i)
      \\
      &+
      n
      \int \diffint{\vect{r}_i}{}
      \int \diffint{\cone{2}}{}
      u(\vect{r}_\cone{1} - \vect{r}_i)
      \check{G}_\text{c}(\cone{1}, \cone{2})
      u(\vect{r}_\cone{2} - \vect{r}_i)
      \check{G}_\text{c}(\cone{2}, \cone{1'})
      \hat{u}_\text{so}(\vect{r}_\cone{1'} - \vect{r}_i)
      .
    \end{split}
    \end{equation}
    \end{widetext}
    \end{subequations}

    Moreover, to have a closed set of equations and a complete quasiclassical theory, the self-energy $\check{\Sigma}[\check{G}_\text{c}]$ is approximated by the quasiclassical self-energy $\check{\sigma}[\check{g}]$ in Eq.~\eqref{eq:eilenberger} which is then a functional of the quasiclassical Green's function $\check{g}$. Performing impurity average and employing the quasiclassical approximation yield simplified expressions for the various self-energy contributions of Eqs.~\eqref{eq:selfenergy_definition} evaluated at position $\vect{R}$, Fermi momentum $\vect{p}_F$ and energy $\epsilon$:
  \begin{widetext}
    \begin{subequations}
    \label{eq:selfenergy_quasi}
    \begin{align}
    \label{eq:selfenergy_quasi_imp}
      \check{\sigma}_\text{imp}
      &=
      -
      \frac{\im}{2}
      \Big\langle
	\frac{1}{\tau(\vect{p} - \vect{q})}
	\check{g}(\vect{R},\vect{q}, \epsilon)	
      \Big\rangle_\text{F}
      ,
      \\
    \label{eq:selfenergy_quasi_so1_0}
      \check{\sigma}_\text{so}^{(1)}
      &=
      \frac{\tilde{\gamma} p_F}{2}
      \Big\langle
	\frac{1}{\tau(\vect{p} - \vect{q})}
	\big[
	  \tauhat{3} \hat{\vect{\alpha}} \cdot (\hat{\vect{p}} \times \hat{\vect{q}})
	  ,
	  \check{g}(\vect{R},\vect{q}, \epsilon)
	\big]_{-}
      \Big\rangle_\text{F}
      +
      \frac{\im \tilde{\gamma}}{4}
      \Big\langle
	\frac{1}{\tau(\vect{p} - \vect{q})}
	\big[
	  \tauhat{3} \hat{\vect{\alpha}} \times (\hat{\vect{p}} - \hat{\vect{q}})
	  ,
	  \vect{\tilde{\nabla}} \check{g}(\vect{R},\vect{q}, \epsilon)
	\big]_{+}
      \Big\rangle_\text{F}
      ,
      \\
    \label{eq:selfenergy_quasi_so}
      \check{\sigma}_\text{so}
      &=
      -
      \frac{\im \gamma^2}{2}
      \Big\langle
	\frac{1}{\tau(\vect{p} - \vect{q})}
	\tauhat{3} \hat{\vect{\alpha}} \cdot (\vect{p} \times \vect{q})
	\check{g}(\vect{R},\vect{q}, \epsilon)
	\tauhat{3} \hat{\vect{\alpha}} \cdot (\vect{p} \times \vect{q})
      \Big\rangle_\text{F}
      ,
      \\
    \label{eq:selfenergy_quasi_skew}
      \check{\sigma}_\text{sk}
      &=
      -
      \frac{\im \gamma}{2}
      \Big\langle
	\frac{1}{\tau_\text{sk}(\vect{p}, \vect{q}, \vect{q}')}
	\big(
	  \tauhat{3} \hat{\vect{\alpha}}
	  \cdot
	  (
	    \vect{p} \times \vect{q}
	  )
	  \check{g}(\vect{R},\vect{q}, \epsilon)
	  \check{g}(\vect{R},\vect{q}', \epsilon)
	  -
	  \check{g}(\vect{R},\vect{q}, \epsilon)
	  \check{g}(\vect{R},\vect{q}', \epsilon)
	  \tauhat{3} \hat{\vect{\alpha}}
	  \cdot
	  (
	    \vect{p} \times \vect{q}'
	  )
	\big)
      \Big\rangle_\text{F}
      \\
    \notag  
      &-
      \frac{\im \gamma}{2}
      \Big\langle
	\frac{1}{\tau_\text{sk}(\vect{p}, \vect{q}, \vect{q}')}
	\check{g}(\vect{R},\vect{q}, \epsilon)
	\tauhat{3} \hat{\vect{\alpha}}
	\cdot
	(
	  \vect{q} \times \vect{q}'
	)
	\check{g}(\vect{R},\vect{q}', \epsilon)
      \Big\rangle_\text{F}
      ,
    \end{align}
    \end{subequations}
  \end{widetext}  
    where $\langle \dots \rangle_\text{F}$ denotes an angular average with respect to $\vect{q}$ (and $\vect{q}'$) at the Fermi surface. We changed the notation of the self-energy in Eq.~\eqref{eq:selfenergy_quasi} by switching from $\check{\Sigma}$ to $\check{\sigma}$ to emphasize that we use the quasiclassical approximation of Eq.~\eqref{eq:selfenergy_definition}. The elastic scattering rate is
    \begin{subequations}
    \label{eq:selfenergy_tau}
    \begin{equation}
    \label{eq:selfenergy_tauimp}
      \frac{1}{\tau(\vect{p} - \vect{q})}
      =
      2 \pi n N_0 \big| u(\vect{p} - \vect{q}) \big|^2
      ,
    \end{equation}
    where $N_0$ is the density of states at the Fermi level in the normal state. 
    The skew scattering rate is
    \begin{equation}
    \label{eq:selfenergy_tauskew}
      \frac{1}{\tau_\text{sk}(\vect{p}, \vect{q}, \vect{q}')}
      =
      2 \pi^2 n N_0^2
      u(\vect{p} - \vect{q}) u(\vect{q} - \vect{q}') u(\vect{q}' - \vect{p})
      .
    \end{equation}
    \end{subequations}
    The skew scattering rate~\eqref{eq:selfenergy_tauskew} is on the order of $1/(N_0 u)$ smaller than the elastic scattering rate~\eqref{eq:selfenergy_tauimp}. A detailed derivation of Eqs.~\eqref{eq:selfenergy_quasi} and~\eqref{eq:selfenergy_tau} is presented in Appendix~\ref{sec:selfenergy}.
    
    
    \subsection{Diffusion Limit}
    \label{subsec:diffusion_limit}
    
    Since elastic impurity scattering is assumed to be strong (dirty limit), the quasiclassical Green's function becomes almost isotropic and we can use an expansion in spherical harmonics up to the first order,
    \begin{equation}
    \label{eq:greens_harmonics}
      \check{g}(\vect{R},\vect{p}_\text{F}, \epsilon)
      \approx
      \check{g}_\text{s}(\vect{R},\epsilon)
      +
      \vect{e}_p \cdot \check{\vect{g}}(\vect{R},\epsilon)
      ,
    \end{equation}
    where $\check{g}_\text{s}$ and $\check{\vect{g}}$ are the isotropic and anisotropic Green's functions, respectively, and $\vect{e}_p = \vect{p}_\text{F}/|\vect{p}_\text{F}|$. Expanding the normalization condition~\eqref{eq:normalization} to the first order yields the useful relations
    \begin{equation}
    \label{eq:normalization_harmonics}
      \check{g}_\text{s}^2
      =
      1
      ,
      \quad
      \big[
	\check{g}_\text{s}, \check{\vect{g}}
      \big]_{+}
      =
      0
      .
    \end{equation}
    We use this expansion (\ref{eq:greens_harmonics}) in the self-energy contributions of Eq.~\eqref{eq:selfenergy_quasi} and retain only the dominant terms. As a conventional example, consider the elastic impurity scattering. Inserting the expansion of the Green's function in spherical harmonics of Eq.~\eqref{eq:greens_harmonics} into the quasiclassical elastic impurity scattering self-energy~\eqref{eq:selfenergy_quasi_imp} and performing the angular average provides
    \begin{subequations}
    \label{eq:selfenergy}
    \begin{equation}
    \label{eq:selfenergy_imp}
    \begin{split}
      \check{\sigma}_\text{imp}(\vect{p}_\text{F})
      &\approx
      -\frac{\im}{2}
      \Big\langle
	\frac{1}{\tau(\vect{p} - \vect{q})}
	(\check{g}_\text{s} + \vect{e}_q \cdot \vect{\check{g}})
      \Big\rangle_\text{F}
      \\
      &=
      -
      \frac{\im}{2 \tau}
      \check{g}_\text{s}
      -
      \frac{\im}{2}
      \Big(
	\frac{1}{\tau}
	-
	\frac{1}{\tau_\text{tr}}
      \Big)
      (
	\vect{e}_p
	\cdot
	\vect{\check{g}}
      )
      ,
    \end{split}
    \end{equation}
    where we dropped the center coordinate $\vect{R}$ and the energy $\epsilon$ for brevity, and
    \begin{equation*}
      \frac{1}{\tau}
      =
      \Big\langle
	\frac{1}{\tau(\vect{p} - \vect{q})}
      \Big\rangle_\text{F}
    \end{equation*}
    is the average elastic scattering rate and
    \begin{equation*}
      \frac{1}{\tau_\text{tr}}
      =
      \Big\langle
	\frac{1}{\tau(\vect{p} - \vect{q})}
	(
	  1
	  -
	  \vect{e}_p
	  \cdot
	  \vect{e}_q
	)
      \Big\rangle_\text{F}
    \end{equation*}
    is the inverse transport relaxation time.
    
    Similarly, we insert the Green's function's expansion (\ref{eq:greens_harmonics}) into the remaining self-energy contributions and perform the angular average. We make use of the fact that $\check{\vect{g}} \ll \check{g}_\text{s}$ and only retain the dominant contributions. To the leading order in the spin-orbit coupling, we then obtain the spin-swapping (``sw''), side-jump (``sj''), and spin-orbit-induced relaxation (``so'') contributions to the self-energy:
    \begin{align}
      \label{eq:selfenergy_sw}
	\check{\sigma}_\text{sw}(\vect{p}_\text{F})
	&
	=
	- \frac{\eta}{3 \tau_\text{sw}}
	\vect{e}_p \cdot
	\big[
	  \tauhat{3} \hat{\vect{\alpha}}
	  \stackrel{\times}{,}
	  \check{\vect{g}}
	\big]_{+}
	,
	\\
      \label{eq:selfenergy_sj}
	\check{\sigma}_\text{sj}(\vect{p}_\text{F})
	&
	=
	- \frac{\im \tilde{\gamma} }{4 \tau_\text{tr}}
	\vect{e}_p \cdot
	\big[
	  \tauhat{3} \hat{\vect{\alpha}}
	  \stackrel{\times}{,}
	  \tilde{\vect{\nabla}} \check{g}_\text{s}
	\big]_{-}
	,
	\\
	  \label{eq:selfenergy_so}
	\check{\sigma}_\text{so}(\vect{p}_\text{F})
	&
	=
	- \frac{3 \im}{16 \tau_\text{so}}
	\big( \tauhat{3} \hat{\vect{\alpha}} \times \vect{e}_p \big)
	\check{g}_\text{s}
	\big( \tauhat{3} \hat{\vect{\alpha}} \times \vect{e}_p \big)
	,
      \intertext{and, using Eq.~\eqref{eq:normalization_harmonics}, we find the skew-scattering contribution}
      \label{eq:selfenergy_sk}
	\check{\sigma}_\text{sk}(\vect{p}_\text{F})
	&
	=
	-
	\frac{\im \eta}{3 \tau_\text{sk}}
	\vect{e}_p
	\cdot
	\big[
	  \tauhat{3} \hat{\vect{\alpha}}
	  \stackrel{\times}{,}
	  \check{g}_\text{s}
	  \check{\vect{g}}
	\big]_{-}
	,
      \end{align}
    \end{subequations}
    where $[ \vect{a} \stackrel{\times}{,} \vect{b} ]_{\pm} = \vect{a} \times \vect{b} \pm \vect{b} \times \vect{a}$ and, again, we omit the arguments $\vect{R}$ and $\epsilon$ for brevity. When evaluating the self-energy to the first order in the spin-orbit coupling \eqref{eq:selfenergy_definition_so1}, a contribution to the side-jump mechanism~\eqref{eq:selfenergy_sj} is also obtained; see Appendices~\ref{subsec:selfenergy_swapping} and~\ref{sec:sidejump}. The elastic scattering rate is
    \begin{subequations}
    \label{eq:selfenergy_scatteringtimes}
    \begin{equation}
    \label{eq:selfenergy_scatteringtimes_imp}
      \frac{1}{\tau}
      =
      2 \pi n N_0
      \Big\langle
	\big|u(\vect{e}_p - \vect{e}_q)\big|^2
      \Big\rangle_\text{F}
      ,
    \end{equation}
    and the inverse transport relaxation time is
    \begin{equation}
    \label{eq:selfenergy_scatteringtimes_imptr}
      \frac{1}{\tau_\text{tr}}
      =
      2 \pi n N_0
      \Big\langle
	\big|u(\vect{e}_p - \vect{e}_q)\big|^2
	(
	  1
	  -
	  \vect{e}_p
	  \cdot
	  \vect{e}_q
	)   
      \Big\rangle_\text{F}
      .
    \end{equation}
    Spin relaxation is determined by the spin-flip scattering rate due to magnetic impurities,
    \begin{equation}
    \label{eq:selfenergy_scatteringtimes_m}
      \frac{1}{\tau_\text{m}}
      =
      \frac{8}{3} \pi n_\text{m} N_0 S (S + 1)
      \Big\langle
	\big|u_\text{m}(\vect{e}_p - \vect{e}_q)\big|^2
      \Big\rangle_\text{F}
      ,
    \end{equation}
    where $S$ is the impurity spin quantum number, and the spin-flip scattering rate due to spin-orbit coupling,
    \begin{equation}
    \label{eq:selfenergy_scatteringtimes_so}
      \frac{1}{\tau_\text{so}}
      =
      \frac{8}{3} \pi \tilde{\gamma}^2 p_\text{F}^2 n N_0
      \Big\langle
	\big|u(\vect{e}_p - \vect{e}_q)\big|^2
	(\vect{e}_p \times \vect{e}_q)^2
      \Big\rangle_\text{F}
      .
    \end{equation}
    The spin-swapping scattering rate is
    \begin{equation}
    \label{eq:selfenergy_scatteringtimes_sw}
      \frac{1}{\tau_\text{sw}}
      =
      \frac{9}{8 \tilde{\gamma}^2 p_\text{F}^2}
      \frac{1}{\tau_\text{so}}
      ,
    \end{equation}
    and the skew scattering rate is
    \begin{equation}
    \begin{split}
    \label{eq:selfenergy_scatteringtimes_sk}
      \frac{1}{\tau_\text{sk}}
      =
      2 \pi^2 n N_0^2 u^3_0
      .
    \end{split}
    \end{equation}
    \end{subequations}
    The results in Eqs.~\eqref{eq:selfenergy} and~\eqref{eq:selfenergy_scatteringtimes} are valid irrespective of the possible anisotropy of the scattering potential, with the exception that the expression for the self-energy due to skew scattering~\eqref{eq:selfenergy_sk} is included only to the lowest order in the anisotropy of the scattering potentials to keep the result compact and simple.\footnote[3]{Expanding the scattering potentials in spherical harmonics, we obtain Eqs.~\eqref{eq:selfenergy_sk} and~\eqref{eq:selfenergy_scatteringtimes_sk} to zeroth order and the first order corrections vanish. This indicates that the anisotropy of the scattering potentials is less relevant for skew scattering.}
    
    The isotropic Green's function $\check{g}_\text{s}$ and the anisotropic Green's function $\check{\vect{g}}$ are obtained using the expansion in spherical harmonics~\eqref{eq:greens_harmonics} and splitting the Eilenberger equation~\eqref{eq:eilenberger} into an even and an odd part with respect to $\vect{e}_p$. To the first order in the spin-orbit coupling strength, the odd part is
    \begin{equation}
    \label{eq:usadel_odd}
    \begin{split}
      0
      &=
      \im v_\text{F} \vect{\tilde{\nabla}} \check{g}_\text{s}
     +
      \frac{\im}{\tau_\text{tr}} \check{g}_\text{s} \check{\vect{g}}+
      \big[
	\tauhat{3} \epsilon, \check{\vect{g}}
      \big]_{-}
      +
      \big[
	\hat{\Delta}, \check{\vect{g}}
      \big]_{-}
      \\
      &-
      \frac{\eta}{3 \tau_\text{sw}}
      \check{g}_\text{s} \Big[
	\big[
	  \tauhat{3} \hat{\vect{\alpha}}
	  ,
	  \check{g}_\text{s}
	\big]_{+}  
	\stackrel{\times}{,}
	\check{g}_\text{s}
	\check{\vect{g}}
      \Big]_{-}
      \\
      &+
      \frac{\im \tilde{\gamma}}{4 \tau_\text{tr}}\check{g}_s\Big[
		\check{g}_s \big[\tauhat{3}\hat{\vect{\alpha}}, \check{g}_s\big]_+ 
		\stackrel{\times}{,}
		\tilde{\nabla}\check{g}_s
		\Big]_-
      \\
      &-
      \frac{\im \eta}{3 \tau_\text{sk}}
      \Big[
	\big[
	  \tauhat{3} \hat{\vect{\alpha}}
	  ,
	  \check{g}_\text{s}
	\big]_{+}
	\stackrel{\times}{,}
	\check{g}_\text{s}
	\check{\vect{g}}
      \Big]_{+}
      ,
    \end{split}
    \end{equation}
    where we used the normalization condition~\eqref{eq:normalization_harmonics}.
    
     The first line on the r.h.s.\ contains the contributions to the zeroth order in the spin-orbit coupling strength. The remaining terms are the contributions to the first order, which we rewrote in a way that simplifies our further calculations. The anisotropic Green's function can be computed to the zeroth order from the first line. The terms involving the energy $\epsilon$ and the superconducting order parameter $\Delta$ can be neglected compared to the dominating contribution arising from the elastic impurity self-energy~\eqref{eq:selfenergy_imp}, and we obtain the following well-known result:
    \begin{equation}
    \label{eq:usadel_anisotropic}
      \check{\vect{g}}^{(0)}
      =
      -
      l_\text{tr} \check{g}_\text{s} \tilde{\vect{\nabla}} \check{g}_\text{s}
      ,
    \end{equation}
    where $l_\text{tr} = v_\text{F} \tau_\text{tr}$ is the impurity mean free path.
    
    By using $\check{\vect{g}} = \check{\vect{g}}^{(0)} + \delta\check{\vect{g}}$ in Eq.~\eqref{eq:usadel_odd} and multiplying by $\im \tau_\text{tr} \check{g}_\text{s}$ from the left, the first-order corrections $\delta\check{\vect{g}}$ to the anisotropic Green's function stemming from spin-orbit coupling are additive,
    \begin{equation*}
      \delta\check{\vect{g}}
      =
      \delta\check{\vect{g}}^\text{(sw)}
      +
      \delta \check{\vect{g}}^\text{(sj)}
      +
      \delta \check{\vect{g}}^\text{(sk)}
      ,
    \end{equation*}
    and are readily obtained by using Eq.~\eqref{eq:normalization_harmonics}. The spin-swapping self-energy~\eqref{eq:selfenergy_sw} contributes with
    \begin{subequations}
    \label{eq:usadel_anisotropic_corrections}
    \begin{gather}
    \label{eq:usadel_anisotropic_corrections_sw}
      \delta\check{\vect{g}}^\text{(sw)}
      =
      \frac{\im \eta l_\text{tr}}{3} \frac{\tau_\text{tr}}{\tau_\text{sw}}
      \Big[
	\big[
	  \tauhat{3} \hat{\vect{\alpha}}
	  ,
	  \check{g}_\text{s}
	\big]_{+}
	\stackrel{\times}{,}
	\tilde{\vect{\nabla}} \check{g}_\text{s}
      \Big]_{-}
      ,
    \intertext{and the correction due to the self-energy contribution to the side-jump mechanism~\eqref{eq:selfenergy_sj} is}
    \label{eq:usadel_anisotropic_correction_sj}
      \delta\check{\vect{g}}^\text{(sj)}
      =
      \frac{\tilde{\gamma}}{4}
      \Big[
	\check{g}_\text{s}
	\big[ \check{g}_\text{s}
	  \tauhat{3} \hat{\vect{\alpha}}
	  ,
	  \check{g}_\text{s}
	\big]_{+}
	\stackrel{\times}{,}
	\tilde{\vect{\nabla}} \check{g}_\text{s}
      \Big]_{-}
      .
    \intertext{Lastly, the correction from the skew scattering self-energy~\eqref{eq:selfenergy_sk} reads as}
    \label{eq:usadel_anisotropic_correction_sk}
      \delta\check{\vect{g}}^\text{(sk)}
      =
      -
      \frac{\eta l_\text{tr}}{3}
      \frac{\tau_\text{tr}}{\tau_\text{sk}}
		\check{g}_\text{s}      
      \Big[
	\big[
	  \tauhat{3} \hat{\vect{\alpha}}
	  ,
	  \check{g}_\text{s}
	\big]_{+}
	\stackrel{\times}{,}
	\tilde{\vect{\nabla}} \check{g}_\text{s}
      \Big]_{+}
      .
    \end{gather}
    \end{subequations}
    We will see that the side-jump mechanism~\eqref{eq:usadel_anisotropic_correction_sj} and skew scattering~\eqref{eq:usadel_anisotropic_correction_sk} both contribute to the same effect, namely, the spin Hall effect. 
    
    Using Eq.~\eqref{eq:usadel_anisotropic} in the part of the Eilenberger equation~\eqref{eq:eilenberger} that is even with respect to $\vect{e}_p$, we obtain the Usadel equation\footnote[4]{The corrections due to spin-orbit coupling vanish because they are proportional to $\epsilon_{ijk} \nabla_i \nabla_k$.}
    \begin{equation}
    \label{eq:usadel}
    \begin{split}
      D \tilde{\vect{\nabla}} \cdot \big( \check{g}_\text{s} \tilde{\vect{\nabla}} \check{g}_\text{s} \big)
      &=
      -
      \im
      \big[
	\tauhat{3} \epsilon, \check{g}_\text{s}
      \big]_{-}
      -
      \im
      \big[
	\hat{\Delta}, \check{g}_\text{s}
      \big]_{-}
      \\
      &+
      \frac{1}{8 \tau_\text{so}}
      \big[
	\hat{\vect{\alpha}} \tauhat{3} \check{g}_\text{s} \tauhat{3} \hat{\vect{\alpha}}, \check{g}_\text{s}
      \big]_{-}
      +
      \frac{1}{8 \tau_\text{m}}
      \big[
	\hat{\vect{\alpha}} \check{g}_\text{s} \hat{\vect{\alpha}}, \check{g}_\text{s}
      \big]_{-}
      ,
      \\
      & \equiv U_{iso}
    \end{split}
    \end{equation}
    where we, at this stage, have included the well-known effect of magnetic impurities causing pair breaking and spin relaxation represented by the scattering lifetime $\tau_m$. We also introduce $D = v_\text{F} l_\text{tr}/3$ as the diffusion constant. Eq.\ (\ref{eq:usadel}) is a counterpart of the drift-diffusion equation and determines the isotropic Green's function.
    
    
    It is possible to obtain simplified scalar equations from the Usadel equation \eqref{eq:usadel}. First, we note that the normalization condition allows the isotropic kinetic Green's function to be represented in terms of the distribution matrix $\hat{h}$,\cite{Rammer2007}
    \begin{equation}
    \label{eq:greens_kinetic_parameterization}
      \hat{g}_\text{s}^\text{K}(\vect{R},\epsilon)
      =
      \hat{g}_\text{s}^\text{R} \hat{h} - \hat{h} \hat{g}_\text{s}^\text{A}
      ,
    \end{equation}
    where the advanced Green's function can be expressed in terms of the retarded Green's function, $\hat{g}_\text{s}^\text{A} = -(\tauhat{3} \hat{g}_\text{s}^\text{R} \tauhat{3})^\dagger$. We assume that the distribution matrix is diagonal with respect to particle-hole space and decompose it according to
    \begin{equation}
    \label{eq:distmatrix}
      \hat{h}
      =
      h^\epsilon + \hat{\alpha}_j \, h^{\epsilon s}_j
      +
      \tauhat{3} (h + \hat{\alpha}_j \, h^s_j)
      \, ,  
    \end{equation}
    where $h_\epsilon$ and $h$ are the energy and particle distribution functions, respectively, and ${h_\epsilon}_j$ and $h_j$ are the spin-energy and spin distribution functions, respectively, where the subscript (j) denotes the spin polarization direction.

In equilibrium, all the distribution functions except the energy distribution, $h^\epsilon$, vanish \citep{morten2004spin}. At equilibrium, the Keldysh function can then be expressed in terms of the retarded and advanced functions in a simple manner

    \begin{equation}
    \label{eq:greens_kinetic_parameterization_eq}
      \hat{g}_\text{s}^\text{K}(\vect{R},\epsilon)
      =
      h_{eq}
      \big(
      	\hat{g}_\text{s}^\text{R} - \hat{g}_\text{s}^\text{A}
      \big)
      ,
    \end{equation}
    where $h_{eq} = \tanh(\epsilon/2T)$.
    
     In general, $g_\text{s}^\text{R}$ and $f_\text{s}^\text{R}$ depend on position and energy and determine how the various transport mechanisms renormalize below the superconducting critical temperature. They are solved by using the retarded part of Eq.~\eqref{eq:usadel} together with the normalization condition $(g_\text{s}^\text{R})^2 - (f_\text{s}^\text{R})^2 = 1$. For energies far above the gap, the functions approach their high-temperature limits ($g_\text{s}^\text{R} \rightarrow 1$ and $f_\text{s}^\text{R} \rightarrow 0$) while they diverge for energies close to the superconducting-induced energy gap. The presence of magnetic impurities in the system suppresses superconductivity and reduces the gap in the energy spectrum.

\subsection{Current and Densities}  
  \label{subsec:currents_and_densities}

Let us now turn to derive expressions that describe the physical particle and spin currents and equations that determine the distribution matrix. We begin by defining a particle density, $n^{(P)}(\cone{1})$, which counts the number of electrons,
  \begin{subequations}
   \label{eq:density_definition}
       \begin{equation}
       \label{eq:density_definition_particle}
	       n^{(P)}(\cone{1})
	       =
	       \frac{1}{2}
	       \lim_{\cone{1'} \to \cone{1}}
	       \operatorname{Tr}
	       \Big[
		       \hat{n}(\cone{1,1'})
	       \Big]
	       =\sum_{\sigma=\uparrow,\downarrow}
				 \langle\psi^\dagger_\sigma(\cone{1}) \psi_\sigma(\cone{1})\rangle \, .
        \end{equation}
    Analogously, we define a spin density, $n^{(S)}(\cone{1})$; a particle energy density; $n^{(P,E)}(\cone{1})$ and a spin energy density, $n^{(S,E)}(\cone{1})$ 
  \begin{equation}
     \label{eq:density_definition_spin}
     	n^{(S)}(\cone{1})
	    =
	    \frac{1}{2}
	    \lim_{\cone{1'} \to \cone{1}}
	    \operatorname{Tr}
      \Big[
		      \vect{\alpha}\,\hat{n}(\cone{1,1'})
	    \Big],
  \end{equation}
\begin{equation}
    \label{eq:density_definition_energy_particle}
	  n^{(P,E)}(\cone{1})
	  =
	  \frac{1}{4}
	  \lim_{\cone{1'} \to \cone{1}}
	  \operatorname{Tr}
	  \Big[
	     \left(
	        \im\hbar\hat{\tau}_3\partial_{t_\cone{1}} - \im\hbar\hat{\tau}_3\partial_{t_\cone{1'}}
	     \right)
	     \hat{n}(\cone{1,1'})
	   \Big],
\end{equation}
	\begin{equation}
     \label{eq:density_definition_energy_spin}
	    n^{(S,E)}(\cone{1})
	    =
	    \frac{1}{4}
	    \lim_{\cone{1'} \to \cone{1}}
	    \operatorname{Tr}
	    \Big[
	      \left(
	       \operatorname{i}\hbar\hat{\tau}_3\partial_{t_\cone{1}} - \operatorname{i}\hbar\hat{\tau}_3\partial_{t_\cone{1'}}
	      \right)
	     \vect{\alpha}\, \hat{n}(\cone{1,1'})
	    \Big],
   \end{equation}
 \end{subequations}
  with $\hat{n}(\cone{1,1'})$ being defined by
	\begin{equation}
     \label{eq:definition_nhat}
	    \hat{n}(\cone{1},\cone{1'})
	    =
			-\frac{\operatorname{i}}{2}\hat{G}^K(\cone{1},\cone{1'})
			+\frac{\operatorname{i}}{2}\hat{\tau}_3
			     \left(
					   \hat{G}^R(\cone{1},\cone{1'}) - \hat{G}^A(\cone{1},\cone{1'})
					 \right).
   \end{equation}	
	
The trace is taken over spin~$\otimes$~particle-hole space. From the densities (\ref{eq:density_definition}), we calculate corresponding currents using the equations of continuity. The current expressions are Wigner transformed, and in terms of the quasiclassical Green's functions, we define a current density matrix

%
	    \begin{align}
    \label{eq:currentmatrix}
      \hat{\vect{\jmath}}(\vect{R},\epsilon)
      =&
      \frac{v_\text{F}}{6} \Big(
      \hat{\vect{g}}^\text{K}(\vect{R},\epsilon) - \tauhat{3}\big(\hat{\vect{g}}^\text{R}(\vect{R},\epsilon) - \hat{\vect{g}}^\text{A}(\vect{R},\epsilon)\big)\Big) \nonumber
	\\      
      &-
      \frac{\gamma p_\text{F} l_\text{tr}}{6 \tau_\text{tr}}
      \big[
	\tauhat{3} \hat{\vect{\alpha}}
	\stackrel{\times}{,}
	\vect{\nabla} \hat{g}^\text{K}_\text{s}(\vect{R},\epsilon)
      \big]_{-}
      ,
    \end{align} 
    
       where the second term is the anomalous current contribution that arises from the anomalous velocity, which is explained in detail in Sec. \ref{subsec:sidejump_anomalous}. 
    

    The particle current density and the spin current density in the quasiclassical approximation are found by taking the proper traces of the current density matrix in Eq. \ref{eq:currentmatrix}
    \begin{subequations}
    \label{eq:current_quasi}
    \begin{align}
    \label{eq:current_quasi_particle}
      J_i(\vect{R})
      &=
      \frac{ N_0}{4}
      \int \diffint{\epsilon}{}
      \operatorname{Tr}
      \big[
	\tauhat{3} \hat{\jmath}_i(\vect{R},\epsilon)
      \big]
      \\
    \notag  
      &=
      N_0
      \int \diffint{\epsilon}{}
      j_i(\vect{R},\epsilon)
    \intertext{and}
    \label{eq:current_quasi_spin}
      J_{ij}(\vect{R})
      &=
      -
      \frac{ N_0}{4}
      \int \diffint{\epsilon}{}
      \operatorname{Tr}
      \big[
	\tauhat{3} \hat{\alpha}_j \hat{\jmath}_i(\vect{R},\epsilon)
      \big]
      \\
    \notag  
      &=
      N_0
      \int \diffint{\epsilon}{}
      j_{ij}(\vect{R},\epsilon)
      ,
      \\
          \label{eq:current_quasi_particle_energy}
      J_{\epsilon i}(\vect{R})
      &=
      \frac{  N_0}{4}
      \int \diffint{\epsilon}{}
      \operatorname{Tr}
      \big[ \epsilon~
	 \hat{\jmath}_{i}(\vect{R},\epsilon)
      \big]
      \\
    \notag  
      &=
      N_0
      \int \diffint{\epsilon}{}
      j_{\epsilon i}(\vect{R},\epsilon)
      \\
      \label{eq:current_quasi_spin_energy}
      J_{\epsilon ij}(\vect{R})
      &=
      \frac{ N_0}{4}
      \int \diffint{\epsilon}{} \epsilon
      \operatorname{Tr}
      \big[\epsilon~
	\hat{\alpha}_{j} \hat{\jmath}_i(\vect{R},\epsilon)
      \big]
      \\
    \notag  
      &=
      N_0
      \int \diffint{\epsilon}{} \epsilon
      j_{\epsilon ij}(\vect{R},\epsilon)
      ,
    \end{align}
    \end{subequations}
    respectively. Here, $j_i$ is the particle current and $j_{ij}$ is the spin current, as introduced in Sec.~\ref{sec:transport}. The expressions for the energy and the spin-energy current densities are derived similarly, but are defined compared to some equilibrium value\citep{silaev2015long}. The second term on the r.h.s.\ of Eq.~\eqref{eq:currentmatrix} results from the anomalous velocity corrections of Eq.~\eqref{eq:velocity_so} and contributes to the side-jump effect; see Appendix~\ref{sec:sidejump}.

    We can now express the gradient of the current using the Usadel equation \eqref{eq:usadel} in terms of the divergence of the matrix current $\hat{\vect{\jmath}}$. Using Eqs.~\eqref{eq:usadel_anisotropic}, \eqref{eq:usadel}, and~\eqref{eq:currentmatrix}, we find\footnotemark[4]
    \begin{equation}
    \label{eq:usadel_kinetic}
    \begin{split}
      \tilde{\vect{\nabla}} \cdot \hat{\vect{\jmath}}
      &=-\frac{1}{2}
      \Big(
      (U_{iso})^K - \tauhat{3}
      \big[
      	(U_{iso})^R - (U_{iso})^A
      \big]
      \Big)
    \end{split}
    \end{equation}
    where the contributions arise from the respective matrix block in Eq. \eqref{eq:usadel}. From Eq.~\eqref{eq:usadel_kinetic}, the diffusion Eq.~\eqref{eq:diffusion} can be derived in terms of the distribution functions in Eq.~\eqref{eq:distmatrix}. The currents in Subsec.~\ref{subsec:transport_currents} are defined {as indicated by Eq.~\eqref{eq:current_quasi} and are calculated using Eqs.~\eqref{eq:usadel_anisotropic} and~\eqref{eq:usadel_anisotropic_corrections} in Eq.~\eqref{eq:currentmatrix}.
    
    The renormalization factors are determined by the components of the retarded/advanced Green's function, where we have inserted the parametrization explained in Appendix \ref{sec:parametrization}, as well as the self-consistency expression. For gap scattering and spin relaxation, they read as
    \begin{subequations}
    \label{eq:renormalizationfactors_1}
    \begin{align}
    \label{eq:renormalizationfactors_alpha}
      \alpha
      &
      =
      2
      \operatorname{Im}[\sinh \theta]
      \operatorname{Re}[e^{-\im \chi} \Delta]
      ,
      \\
      \label{eq:renormalizationfactors_alphaE}
      \alpha^\epsilon
      &
      =
      2
      \operatorname{Re}[\sinh \theta]
      \operatorname{Im}[e^{-\im \chi} \Delta]
      \\
    \label{eq:renormalizationfactors_asoTS}
      \alpha_\text{so}
      &
      =
      \operatorname{Re}[ \cosh \theta ]^2 - \operatorname{Re}[ \sinh \theta]^2
      ,
      \\
          \label{eq:renormalizationfactors_asoLS}
      \alpha^\epsilon_\text{so}
      &
      =
		\operatorname{Re}[ \cosh \theta ]^2 + \operatorname{Im}[ \sinh \theta]^2      		,
      \\
    \label{eq:renormalizationfactors_amTS}
      \alpha_\text{m}
      &
      =
      \operatorname{Re}[ \cosh \theta ]^2 + \operatorname{Re}[ \sinh \theta]^2
      ,
      \\
    \label{eq:renormalizationfactors_amLS}
      {\alpha}^\epsilon_\text{m}
      &
      =
\operatorname{Re}[ \cosh \theta ]^2 - \operatorname{Im}[ \sinh \theta]^2      ,
    \end{align}
    \end{subequations}

    The renormalized diffusion constants are
    \begin{subequations}
    \label{eq:renormalizationfactors_2}
    \begin{align}   
    \label{eq:renormalizationfactors_DL}
      D_\epsilon
      &
      =
      \frac{D}{2}
	\big(      
      1 + |\cosh\theta|^2 - |\sinh\theta|^2
      \big)
      ,
      \\
    \label{eq:renormalizationfactors_DT}
      D_\text{p}
      &
      =
      \frac{D}{2}
      \big(
      1 + |\cosh\theta|^2 + |\sinh\theta|^2
      \big)
      ,
    \end{align}
    \end{subequations}
    and
    \begin{equation}
    \label{eq:renormalizationfactors_NS}
      N_\text{S}
      =
      \operatorname{Re}[\cosh\theta ]
    \end{equation}
    is the density of states in the superconductor normalized by the density of states in the normal state. We also define the following currents related to the supercurrents in the system

\begin{subequations}
\begin{equation}
	\vect{v}_s = \nabla \chi - 2 \frac{e}{\hbar}\vect{A}_i
\end{equation}
    \begin{equation}
    \label{eq:normal_supercurrent}
    	j^{sc}_i = \big\{2 \operatorname{Im}(\sinh^2\theta) \big\}
    	v_{si}
    \end{equation}
    \begin{equation}
    	j^{sc,2}_i = \big\{1 - |\cosh\theta|^2 + |\sinh \theta|^2 \big\}
    	v_{si}
    \end{equation}
    \begin{equation}
    	\vect{\mathcal{R}}^p_i =
    	-2 \operatorname{Im}
    	\big(
    	\sinh \theta
    	\big)
    	\nabla_i
    	(\operatorname{Re}\theta)
    \end{equation}
        \begin{equation}
    	\vect{\mathcal{R}}^\epsilon_i =
    	-2 \operatorname{Re}
    	\big(
    	\sinh \theta
    	\big)
    	\nabla_i
    	(\operatorname{Im}\theta)
    \end{equation}
\end{subequations}    

This completes our derivations of the diffusion equations and the associated relations between the currents and the spatial variations of the densities.

%
%
    
  \section{Conclusion}  
  \label{sec:conclusion}      

    We have derived diffusion equations for the transport of spin, particle, and energy in the elastic transport regime,
    including scattering from magnetic and non-magnetic impurities and from spin-orbit coupling.
    We find that the spin Hall angle is renormalized by the reduced density of states in the superconducting state. However, the spin-swapping coefficient does not explicitly depend on the superconducting correlations but rather is influenced by the superconducting state through the renormalized diffusion coefficients.

    In a two-dimensional geometry, we find a large enhancement of the spin-swapping effects. This result implies that it should be possible to measure the influence of superconductivity on these largely unexplored transport properties.
    
    We thank Jacob Linder for stimulating discussions.

\appendix

%
%
    %
  \section{Fourier Transform}
  \label{sec:fourier}
    We define the Fourier transforms as
    \begin{subequations}
    \label{eq:Fourier}
    \begin{align}
      x(\vect{r}, t)
      &=
      \int \! \frac{\mathrm{d}\vect{q}}{(2 \pi \hbar)^3}\,
      e^{-i \vect{q} \vect{r}/\hbar}
      \int \frac{\!\mathrm{d}\epsilon \!}{2 \pi \hbar}\,
      e^{ i \epsilon t /\hbar}
      x_F(\vect{q}, \epsilon)
      ,
      \\
      x_F(\vect{q}, \epsilon)
      &=
      \int \diffint{\vect{r}}{}
      e^{i \vect{q} \vect{r} /\hbar}
      \int \diffint{t}{}
      e^{-i \epsilon t /\hbar}
      x(\vect{r}, t)
      ,
    \end{align}
    \end{subequations}
    where the subscript "F" indicates that we are referring to the Fourier transform and not the Wigner transform. 
%

%
%
    %
  \section{Wigner Transform}
  \label{sec:covariantwigner}

	In this section, we will introduce the Wigner transform which we will use extensively. We follow the conventions in Ref.~[\onlinecite{konschelle2014transport}] for an Abelian and spin-independent vector field.

 We can relate the Fourier transform and the Wigner transform by using a translation operator. For a function that depends on the space-times $\cone{1} = X + z/2$ and $\cone{1'} = X -z/2$, where $X$ and $z$ are the absolute and relative space-times, we have

\begin{equation}
x_F (p,X) = \int dz~e^{-ipz/\hbar} \left[e^{\frac{z}{2} \partial_X}x(X,X')e^{-\frac{z}{2} \partial'_{X'}}\right]\Biggr|_{X=X'},
\end{equation}

where $z \equiv (t,\vect{r})$, $X \equiv (T,\vect{R})$, and $p = (\epsilon,\vect{p})$. The inner product is defined according to the mostly minus metric, as outlined in Sec.~\ref{sec:transport}. To arrive at a co-variant transform, we let $\partial_{X^\mu} \rightarrow D_\mu(X)$ and $\partial'_{X^\mu} \rightarrow D'_\mu(X)$, where the co-variant derivative is defined according to Eq.~\eqref{eq:covariant_derivative}, which results in

\begin{equation}
\label{eq:wignerco}
\begin{split}
x (p,X) &= \int dz~e^{-ipz/\hbar} \\
&\times
\left[\text{exp}\left(\frac{z^\mu}{2} D_\mu \right) x(X,X')\text{exp}\left(-\frac{z^\mu}{2} D'_\mu \right)\right]\Biggr|_{X=X'},
\end{split}
\end{equation}

which is how we define the co-variant Wigner transform. Note that we write the co-variant Wigner transform \textit{without} a subscript "$F$". We define the \textit{connector}, $\hat{U}$,

\begin{equation}
	\hat{U} (b,a) \equiv \exp\left[\im \tauhat{3} (b-a)^\mu \int_0^1 ds A_\mu\left(a + (b-a)s\right)\right], 
\end{equation}

and to ease notation, we also define $\hat{U}_1 = \hat{U}(\cone{1}, X)$ and $\hat{U}_2 = \hat{U}(X, \cone{1'})$. In terms of the connectors, the Wigner transform becomes

\begin{align}
x(p,X) &= \int dz \left[e^{-ipz/\hbar}\hat{U}_1~
 x(\cone{1},\cone{1'})~\hat{U}_2. \right]\nonumber
\label{eq:wignerco-connector}
\end{align}
The inverse transform is
\begin{equation}
x(\cone{1},\cone{1'}) = \int \frac{dp}{(2\pi)^4} \left[e^{\im pz/\hbar}\hat{U}_1^\dagger ~x(p,X)~\hat{U}_2^\dagger \right].
\label{eq:wignerco-connector2}
\end{equation}

Our Green's functions are matrices in electron-hole space. This carries over to a matrix structure in the connector. In electron-hole space, the Wigner transform of our Green function is defined as

\begin{equation}
	\hat{G}(p,X) = \int dz~e^{-\im pz/\hbar} \left[\hat{U}_1
 \hat{G}(\cone{1},\cone{1'})\hat{U}_2. \right],
\end{equation}

 In the following, we will expand the connectors in the gradient approximation. 

%
%

  \section{Parametrization}  
  \label{sec:parametrization}
  
  To simplify the calculations, we apply the $\theta$-parametrization. The retarded Green's function is
  
\begin{equation}
	\hat{g}^R (\epsilon)
	=
	\begin{pmatrix}
		\hat{1}\cosh\left[\theta(\epsilon) \right] & \im \sigma_2 \sinh\left[\theta(\epsilon) \right] e^{i \chi(\epsilon)} \\
		\im \sigma_2 \sinh\left[\theta(\epsilon) \right] e^{-i \chi(\epsilon)} & -\hat{1}\cosh\left[\theta(\epsilon) \right]
	\end{pmatrix}.	
\end{equation}

The advanced function can be found using the relation $	\hat{g}^A (\textbf{R},\epsilon) = - \left(\tau_ 3 \hat{g}^R (\textbf{R},\epsilon) \tau_3\right)^\dagger$. Inspecting the elements of the retarded Green's function in Eq.~\eqref{eq:greens_definition_R} of the Green's function, and the normalization condition, we obtain the following symmetries for $\theta(\textbf{R},\epsilon)$ and $\chi(\textbf{R},\epsilon)$

\begin{subequations}
\begin{equation}
	\chi(\epsilon) = \chi^* (-\epsilon),
\end{equation}
\begin{equation}
	\theta(\epsilon) = -\theta^*(-\epsilon).
\end{equation}
\end{subequations}

We insert these when we calculate the current.

%
%

  \section{Self-consistency}  
  \label{sec:self_consistency}
  
	The superconducting gap must be calculated self-consistently, and we can express the gap using the Keldysh Green's function in the following way
	
	\begin{equation}
		\Delta(\cone{1}) = -\frac{1}{8}N_0 \lambda \int d\epsilon~
		\text{Tr}\Bigg[
		\frac{(\tauhat{1} - \im\tauhat{2})\alpha_3}{2}
		\hat{g}_s^K(\vect{R},\epsilon)
		\Bigg],
\end{equation}	  

where $\lambda$ is the strength of the pairing potential. Using our ansatz for the distribution function in Eq. \ref{eq:greens_kinetic_parameterization}, we can express the gap in terms of distribution functions and the parametrization parameters

\begin{equation}
	\Delta(\cone{1}) = \frac{N_0 \lambda}{2}
	\int d\epsilon~
e^{\im\chi}	
	\Big[
		- \operatorname{Re}(\sinh \theta) h^\epsilon
		+ \im \operatorname{Im}(\sinh \theta) h
	\Big].
\end{equation}
We use this relation in our expressions. Since we have chosen $\chi$ to be a real number, the phase of the order parameter is also real. 

%
%
    
  \section{The Self-Energy}  
  \label{sec:selfenergy}
  
    Here, we will calculate the contributions to the quasiclassical self-energy~\eqref{eq:selfenergy_quasi} and outline how the self-energy contributions to the Eilenberger equation~\eqref{eq:eilenberger} and the Usadel equations~\eqref{eq:usadel_odd} and~\eqref{eq:usadel} are obtained. We include effects from elastic impurity scattering, magnetic impurities and spin-orbit coupling within the self-consistent Born approximation; see the Feynman diagrams in Fig.~\ref{fig:selfenergy_born_1}. Skew scattering only appears beyond the self-consistent Born approximation to at least the third order in the potential $u$,\cite{Kohn1957, Engel2007}. We take this lowest order contribution to the skew scattering into account; see Fig.~\ref{fig:selfenergy_born_2}. The self-energies caused by elastic scattering, magnetic impurities, and the contribution from spin-orbit scattering to spin relaxation are well known, but we also include their brief derivation here for completeness and consistency in the notation.    
    
%
%
    
  \subsection{Elastic Impurity Scattering}
  \label{subsec:selfenergy_imp}
  
    Using the Fourier representation of the elastic impurity scattering potential $u(\vect{r} - \vect{r}_i)$, its contribution to the self-energy~\eqref{eq:selfenergy_definition_imp} is
    \begin{equation*}
      \check{\Sigma}_\text{imp}(\cone{1}, \cone{1'})
      =
      n
      \int \! \frac{\mathrm{d}\vect{q}}{(2 \pi \hbar )^3}\,
      \big| u(\vect{q}) \big|^2
      \e{-\im \vect{q} \cdot \vect{r}/\hbar}
      \check{G}_\text{c}(\cone{1}, \cone{1'})
      ,
    \end{equation*}
    where $\vect{r} = \vect{r}_\cone{1} - \vect{r}_\cone{1'}$ is the relative position. Next, we Fourier transform the relative spatial and temporal coordinates using the Fourier transform of Eq.~\eqref{eq:Fourier}: 
    \begin{equation}  
    \label{eq:selfenergy_full_imp}
      \check{\Sigma}_\text{imp}(\vect{R},\vect{p}, \epsilon)
      =
      n
      \int \! \frac{\mathrm{d}\vect{q}}{(2 \pi \hbar )^3}\,
      \big| u(\vect{p} - \vect{q}) \big|^2
      \check{G}_\text{c}(\vect{R}, \vect{q}, \epsilon)
      .
    \end{equation}
    Within the quasiclassical framework we can approximate $\int \! \frac{\mathrm{d}\vect{q}}{(2 \pi \hbar)^3} \approx N_0 \int \diffint{\xi_q}{} \int \! \frac{\mathrm{d}\vect{e}_q}{4 \pi}$, where $N_0$ is the density of states and $\vect{e}_q = \vect{q}_\text{F}/|\vect{q}_\text{F}|$, such that the quasiclassical approximation to the self-energy is
    \begin{equation}
    \label{eq:self_energy_imp}
      \check{\sigma}_\text{imp}(\vect{R},\vect{p}_\text{F}, \epsilon)
      =
      -
      \frac{\im}{2}
      \Big\langle
	\frac{1}{\tau(\vect{p} - \vect{q})}
	\check{g}(\vect{R},\vect{q}, \epsilon)	
      \Big\rangle_\text{F}
      ,
    \end{equation}
    where $\langle \dots \rangle_\text{F} = \int \! \frac{\mathrm{d}\vect{e}_q}{4 \pi} \dots$ denotes an angular average over all momentum directions at the Fermi surface and the elastic scattering rate is
    \begin{equation}
      \frac{1}{\tau(\vect{p} - \vect{q})}
      =
      2 \pi n N_0 \big| u(\vect{p} - \vect{q}) \big|^2
      .
    \end{equation}
    We changed the notation of the self-energy in Eq.~\eqref{eq:self_energy_imp} to the symbol $\check{\sigma}$ to reflect that it is the quasiclassical approximation of Eq.~\eqref{eq:selfenergy_full_imp}. This is a standard result for the elastic scattering contribution to the self-energy that we included for completeness.

  \subsection{First Order in Spin-Orbit Coupling}
  \label{subsec:selfenergy_swapping}  
    
    The contributions to the self-energies above are well known. Let us now consider the nontrivial effect of the spin-orbit interaction to the first order in the spin-orbit interaction strength. Inserting the expressions for $\hat{u}_\text{so}$ and $u$ into Eq.~\eqref{eq:selfenergy_definition_so1} yields
    \begin{equation*}
    \begin{split}
      & \;
      \check{\Sigma}_\text{so}^{(1)}(\cone{1}, \cone{1'})
      \\
      =& \;
      - \frac{\gamma n}{\hbar}
      \int \! \frac{\mathrm{d}\vect{q}}{(2 \pi \hbar)^3}\,
      \big|  u(\vect{q}) \big|^2
      \e{-\im \vect{q} \cdot \vect{r}/\hbar}
      \tauhat{3}\hat{\vect{\alpha}}\cdot(\tilde{\vect{\nabla}}_{\vect{r}_{\cone{1}}} \check{G}_\text{c}(\cone{1}, \cone{1'})  \times \vect{q}) 
      \\
      & \;
       -\frac{\gamma n}{\hbar}
      \int \! \frac{\mathrm{d}\vect{q}}{(2 \pi \hbar)^3}\,
      \big|  u(\vect{q}) \big|^2
      \e{-\im \vect{q} \cdot \vect{r}/\hbar}
      (\check{G}_\text{c}(\cone{1}, \cone{1'})\tilde{\vect{\nabla}}'_{\vect{r}_{\cone{1'}}} \times \vect{q}) \cdot  \tauhat{3}\hat{\vect{\alpha}}
      ,
    \end{split}
    \end{equation*}
   where $\tilde{\vect{\nabla}}$ is defined as
	\begin{subequations}	
	\begin{equation}
		\tilde{\vect{\nabla}} X = (\vect{\nabla}X) -\im \frac{e}{\hbar}\vect{A} [\tauhat{3},X]_-.
	\end{equation},
	\begin{equation}
		X\tilde{\vect{\nabla}}  = (X\vect{\stackrel{\leftarrow}{\nabla}}) + \im \frac{e}{\hbar}\vect{A} [X,\tauhat{3}]_-.
	\end{equation}
	\end{subequations} 
    
    In the quasiclassical approximation we obtain from this
    \begin{equation}
    \label{eq:self_energy_firstorder}
    \begin{split}
      \check{\sigma}_\text{so}^{(1)}(\vect{p}_\text{F})
      &=
      \frac{\tilde{\gamma} p_F}{2}
      \Big\langle
	\frac{1}{\tau(\vect{p} - \vect{q})}
	\big[
	   \tauhat{3}\hat{\vect{\alpha}} \cdot (\hat{\vect{p}} \times \hat{\vect{q}})
	  ,
	  \check{g}(\vect{q})
	\big]_{-}
      \Big\rangle_\text{F}
      \\
      &
      + \frac{\im \tilde{\gamma}}{4} 
      \Big\langle
	\frac{1}{\tau(\vect{p} - \vect{q})}
	\big[
	   \tauhat{3}\hat{\vect{\alpha}} \times (\hat{\vect{p}} - \hat{\vect{q}})
	  ,
	  \tilde{\vect{\nabla}} \check{g}(\vect{q})
	\big]_{+}
      \Big\rangle_\text{F}
      ,     
    \end{split}
    \end{equation}
    where we omitted $\vect{R}$ and $\epsilon$ for brevity. We also introduced the dimensionless parameter $\tilde{\gamma} = \gamma p_F /\hbar^2$. 

    The first term on the r.h.s. of Eq. \ref{eq:self_energy_firstorder} gives rise to the spin-swapping effect \cite{Lifshits2009,Sadjina2012}. The second contributes to the side-jump mechanism but is only present when considering the next-to-leading order in the gradient approximation. The side-jump mechanism is discussed in more detail in Appendix~\ref{sec:sidejump}.


%
%

  \subsection{Second Order in Spin-Orbit Coupling}
  \label{subsec:selfenergy_so}  
    
    Similarly, we obtain from Eq.~\eqref{eq:selfenergy_definition_so} to the lowest order in the quasiclassical approximation the self-energy to the second order in the spin-orbit coupling strength:
    \begin{equation}
    \label{eq:selgenergy_second_quasi}
      \check{\sigma}_\text{so}(\vect{p}_\text{F})
      =
      -
      \frac{\im \tilde{\gamma}^2 p_F^2}{2}
      \Big\langle
	\frac{1}{\tau(\vect{p} - \vect{q})}
	\tauhat{3} \hat{\vect{\alpha}} \cdot (\hat{\vect{p}} \times \hat{\vect{q}})
	\check{g}(\vect{q})
	\tauhat{3} \hat{\vect{\alpha}} \cdot (\hat{\vect{p}} \times \hat{\vect{q}})
      \Big\rangle_\text{F}
      ,
    \end{equation}
    where we again omitted $\vect{R}$ and $\epsilon$. This self-energy contribution describes spin-orbit-induced spin relaxation. 
    

%
%
   
  \begin{widetext} 
  \subsection{Skew Scattering}
  \label{subsec:selfenergy_skew}  
  
    We include skew scattering to the lowest order in the gradient approximation. Inserting Eqs.~\eqref{eq:potentials_imp} and~\eqref{eq:potentials_soc} into the skew-scattering contribution to the self-energy~\eqref{eq:selfenergy_definition_skew} provides
	
	    \begin{equation}
	    \label{eq:init_selfenergy_skew}
    \begin{split}
      \check{\Sigma}_\text{sk}(\cone{1}, \cone{1'})
      &=
      -\frac{\gamma n}{\hbar}
      \int \! \frac{\mathrm{d}\vect{q}}{(2 \pi \hbar)^3}\,
      \int \! \frac{\mathrm{d}\vect{q}'}{(2 \pi \hbar)^3}\,
      u(\vect{q}) u(-\vect{q} - \vect{q}') u(\vect{q}')
      \int \diffint{\cone{2}}{}
      \e{-\im \vect{q} \cdot \vect{r}}
      \e{-\im \vect{q}' \cdot (\vect{r}_\cone{2} - \vect{r}_\cone{1'})}
      \\
      &\times
      \Bigg[
		(\tauhat{3}\hat{\vect{\alpha}} \times \vect{q})\cdot\left(\vect{D}(\vect{r}_1)\check{G}_\text{c}(\cone{1}, \cone{2})\right)\check{G}_\text{c}(\cone{2}, \cone{1'}) 
		+ \check{G}_\text{c}(\cone{1}, \cone{2})(\tauhat{3}\hat{\vect{\alpha}} \times \vect{q'})\cdot\left(\vect{D}(\vect{r}_2)\check{G}_\text{c}(\cone{2}, \cone{1'})\right)
      \\
      &+
      \check{G}_\text{c}(\cone{1}, \cone{2}) \left(\check{G}_\text{c}(\cone{2}, \cone{1'})\vect{D}'(\vect{r}_{1'})\right)\cdot(\tauhat{3} \hat{\vect{\alpha}} \times (\vect{q} + \vect{q'}))
      \Bigg]
      ,
    \end{split}
    \end{equation}

    where we performed a partial integration in the Dyson equation in the last term and $\vect{D}'(\vect{r}_{1'})$ acts to the left. We rewrite the Green's functions in terms of their respective center-of-mass and relative coordinates and, for example, use

    \begin{equation*}
      \check{G}_\text{c}(\cone{1}, \cone{2})
      =
      \check{G}_\text{c}\Big(\frac{\vect{r}_\cone{1} + \vect{r}_\cone{2}}{2}, \vect{r}_\cone{1} - \vect{r}_\cone{2}, t_\cone{1} - t_\cone{2}\Big)
      =
      \check{G}_\text{c}\Big(\vect{R} + \frac{\vect{r}_\cone{2} - \vect{r}_\cone{1'}}{2}, \vect{r}/2 - (\vect{r}_\cone{2} - \vect{R}), t_\cone{1} - t_\cone{2}\Big)
      .
    \end{equation*}
    
    We also disregard the correction to the center-of-mass coordinate $\vect{R} = (\vect{r}_\cone{1} + \vect{r}_\cone{1'})/2$ to the lowest order in the gradient approximation, such that
    \begin{equation*}
    \begin{split}
      \check{G}_\text{c}(\cone{1}, \cone{2})
      &\approx
      \check{G}_\text{c}\big(\vect{R}, \vect{r}/2 - (\vect{r}_\cone{2} - \vect{R}), t_\cone{1} - t_\cone{2}\big)
      ,
      \\
      \check{G}_\text{c}(\cone{2}, \cone{1'})
      &\approx
      \check{G}_\text{c}\big(\vect{R}, \vect{r}/2 + (\vect{r}_\cone{2} - \vect{R}), t_\cone{2} -  t_\cone{1'}\big)
      .
    \end{split}  
    \end{equation*}
    After inserting the Wigner coordinates and Fourier transforming Eq. \ref{eq:init_selfenergy_skew}, we have
	
	\begin{equation*}
	\begin{split}
      \check{\Sigma}_\text{sk}(\vect{R}, \vect{p}, \epsilon)
      &=
      -\frac{\gamma n}{\hbar}
      \int \! \frac{\mathrm{d}\vect{q}}{(2 \pi \hbar)^3}\,
      \int \! \frac{\mathrm{d}\vect{q}'}{(2 \pi \hbar)^3}\,
      u(\vect{q}) u(-\vect{q} - \vect{q}') u(\vect{q}')
      \int \diffint{\vect{r}}{}
      \e{ \im \vect{p} \cdot \vect{r}/\hbar}
      \int \diffint{\vect{r}_\cone{2}}{}
      \e{-\im \vect{q} \cdot \vect{r}/\hbar}
      \e{- \im \vect{q}' \cdot \big(\vect{r}_2 - \vect{R} + \vect{r}/2\big)}
      \\
      &\times
      \Bigg[
	      (\tauhat{3}\hat{\vect{\alpha}} \times \vect{q})\cdot\left(\vect{D}(\vect{R} + \vect{r}/2)\check{G}_\text{c}(\vect{R}, \vect{R} + \vect{r}/2 - \vect{r}_2,\epsilon)\right)\check{G}_\text{c}(\vect{R},\vect{r}_2 - \vect{R} + \vect{r}/2,\epsilon)
	      \\
	      &+
	      \check{G}_\text{c}(\vect{R}, \vect{R} + \vect{r}/2 - \vect{r}_2,\epsilon)(\tauhat{3}\hat{\vect{\alpha}} \times \vect{q'})\cdot\left(\vect{D}(\vect{r}_2)\check{G}_\text{c}(\vect{R},\vect{r}_2 - \vect{R} + \vect{r}/2,\epsilon)\right)
	      \\
	      &+
	       \check{G}_\text{c}(\vect{R}, \vect{R} + \vect{r}/2 - \vect{r}_2,\epsilon) \left(\check{G}_\text{c}(\vect{R},\vect{r}_2 - \vect{R} + \vect{r}/2)\vect{D}^\dagger(\vect{R} - \vect{r}/2,\epsilon)\right)\cdot(\tauhat{3}\hat{\vect{\alpha}} \times (\vect{q} + \vect{q'}))	      
      \Bigg]
      \end{split}
	\end{equation*}	    
        
    where we used that, in a stationary case, the convolution with respect to the time variables reduces to a simple product
    \begin{equation*}
      \int \diffint{t}{}
      \e{\im \epsilon t}
      \int \diffint{t_\cone{2}}{}
      \check{G}_\text{c}(t_\cone{1} - t_\cone{2})
      \check{G}_\text{c}(t_\cone{2} -  t_\cone{1'})
      =
      \check{G}_\text{c}(\epsilon)
      \check{G}_\text{c}(\epsilon)
      .
    \end{equation*}
    Next, we introduce new variables according to
    \begin{equation*}
      \vect{r}
      =
      \vect{x} + \vect{y}
      ,
      \quad
      \vect{r_\cone{2}} - \vect{R}
      =
      \frac{\vect{x} - \vect{y}}{2}
      ,
      \quad
      \frac{\partial(\vect{r}, \vect{r_\cone{2}})}{\partial(\vect{x}, \vect{y})}
      =
      -1
      ,
    \end{equation*}
    and consequently, we obtain
    
	\begin{equation*}
    \begin{split}
      \check{\Sigma}_\text{sk}(\vect{R}, \vect{p}, \epsilon)
      &=
      -\frac{\gamma n}{\hbar}
      \int \! \frac{\mathrm{d}\vect{q}}{(2 \pi \hbar)^3}\,
      \int \! \frac{\mathrm{d}\vect{q}'}{(2 \pi \hbar)^3}\,
      u(\vect{q}) u(-\vect{q} - \vect{q}') u(\vect{q}')
      \int \diffint{\vect{x}}{}
      \e{ \im (\vect{p} - \vect{q}) \cdot \vect{x}/\hbar}
      \int \diffint{\vect{y}}{}
      \e{ \im (\vect{p} - \vect{q} - \vect{q'}) \cdot \vect{y}/\hbar}
      \\
      &\times
      \Bigg[
	(\tauhat{3}\hat{\vect{\alpha}} \times \vect{q})\cdot\left(\vect{\partial}_{\vect{x}}\check{G}_\text{c}(\vect{R}, \vect{x},\epsilon)\right)\check{G}_\text{c}(\vect{R},\vect{y},\epsilon)
	+
	\check{G}_\text{c}(\vect{R}, \vect{x},\epsilon)\tauhat{3}(\hat{\vect{\alpha}} \times \vect{q'})\cdot\left((- \vect{\partial}_{\vect{x}} + \vect{\partial}_{\vect{y}})\check{G}_\text{c}(\vect{R},\vect{y},\epsilon)\right)
	\\
	&-
	\check{G}_\text{c}(\vect{R}, \vect{x},\epsilon) \left(\vect{\partial}_{\vect{y}}\check{G}_\text{c}(\vect{R},\vect{y},\epsilon)\right)\cdot(\tauhat{3}\hat{\vect{\alpha}} \times (\vect{q} + \vect{q'}))
      \Bigg]
      ,
    \end{split}
    \end{equation*}    
    
   	where we only retained the lowest-order terms in the quasiclassical approximation, which reduced the co-variant derivative to normal derivatives.
   	
    Performing out the partial integration provides
    \begin{equation}
    \label{eq:selfenergy_skew}
    \begin{split}
      \check{\Sigma}_\text{sk}(\vect{p})
      &=
      \im\frac{\gamma n}{\hbar^2}
      \int \! \frac{\mathrm{d}\vect{q}}{(2 \pi \hbar)^3}\,
      \int \! \frac{\mathrm{d}\vect{q}'}{(2 \pi \hbar)^3}\,
      u(\vect{p} - \vect{q}) u(\vect{q} - \vect{q}') u(\vect{q}' - \vect{p})
      \\
      &\times
      \Big(
	 \tauhat{3}\hat{\vect{\alpha}}	
	\cdot
	(
	  \vect{p} \times \vect{q}
	)
	\check{G}_\text{c}(\vect{q})
	\check{G}_\text{c}(\vect{q}')
	+
	\check{G}_\text{c}(\vect{q})
	 \tauhat{3}\hat{\vect{\alpha}}
	\cdot
	(
	  \vect{q} \times \vect{q}'
	)
	\check{G}_\text{c}(\vect{q}')
	-
	\check{G}_\text{c}(\vect{q})
	\check{G}_\text{c}(\vect{q}')
	 \tauhat{3}\hat{\vect{\alpha}}
	\cdot
	(
	  \vect{p} \times \vect{q}'
	)
      \Big)
      ,
    \end{split}
    \end{equation}
    where we omitted the arguments $\vect{R}$ and $\epsilon$ for brevity. Eq.~\eqref{eq:selfenergy_skew} is in agreement with recent results\cite{Raimondi2012} that are valid for a normal metal only. Our treatment is a generalization to include skew scattering in the superconducting state. In the quasiclassical approximation, the skew-scattering contribution to the self-energy is
    \begin{equation*}
      \check{\sigma}_\text{sk}(\vect{p}_\text{F})
      =
      -
      \frac{\im \tilde{\gamma} p_F}{2}
      \Big\langle
	\frac{1}{\tau_\text{sk}(\vect{p}, \vect{q}, \vect{q}')}
	\big(
	  \tauhat{3}\hat{\vect{\alpha}}
	  \cdot
	  (
	    \hat{\vect{p}} \times \hat{\vect{q}}
	  )
	  \check{g}(\vect{q})
	  \check{g}(\vect{q}')
	  +
	  \check{g}(\vect{q})
	  \tauhat{3}\hat{\vect{\alpha}}
	  \cdot
	  (
	    \hat{\vect{q}} \times \hat{\vect{q}}'
	  )
	  \check{g}(\vect{q}')
	  -
	  \check{g}(\vect{q})
	  \check{g}(\vect{q}')
	  \tauhat{3}\hat{\vect{\alpha}}
	  \cdot
	  (
	    \hat{\vect{p}} \times \hat{\vect{q}}'
	  )
	\big)
      \Big\rangle_\text{F}
      ,
    \end{equation*}
    where
    \begin{equation*}
      \frac{1}{\tau_\text{sk}(\vect{p}, \vect{q}, \vect{q}')}
      =
      2 \pi^2 n N_0^2
      u(\vect{p} - \vect{q}) u(\vect{q} - \vect{q}') u(\vect{q}' - \vect{p})
    \end{equation*}
    is the skew-scattering rate. Note that the skew-scattering rate $1/\tau_\text{sk}$ is a factor on the order of $1/(N_0 u)$ smaller than the elastic scattering rate $1/\tau$.
  
  \end{widetext}



%
%
  
  \section{The Side-Jump Mechanism}  
  \label{sec:sidejump}

    The derivation of the side-jump contribution to the spin Hall effect is a subtle issue\cite{Berger1970, Lyo1972, Nozieres1973, Engel2007, Lee2008} because there are three\footnote[5]{The situation further complicates in the presence of magnetic fields or when the spin operator is time dependent, for example, and even more contributions emerge. To make matters worse, several of these terms cancel while some add up giving rise to factors of $2$. For a detailed discussion see, for example, Ref.~\onlinecite{Nozieres1973}.} contributions to this effect, and one or two of these continue to be overlooked in many works: \romannumeral 1) A contribution arises from the self-energy to the first order in the spin-orbit interaction of Eq.~\eqref{eq:selfenergy_sj}. This contribution only appears beyond the lowest-order gradient approximation and is therefore often disregarded. However, within the quasiclassical approximation, it is of the same order as the other spin-orbit-induced self-energy contributions of Eq.~\eqref{eq:selfenergy} and must be included. It enters in the first term of the matrix current~\eqref{eq:currentmatrix} via the correction to the anisotropic Green's function due to the side-jump self-energy~\eqref{eq:usadel_anisotropic_correction_sj}. \romannumeral 2) Additionally, there is an anomalous current contribution~\eqref{eq:velocity_so_1} from the spin-orbit-induced correction to the velocity operator. \romannumeral 3) Finally, the spin-orbit coupling is expressed in an effective model with a renormalized coupling strength that is typically much larger than the vacuum value. In this effective theory, the position operator also acquires an additional spin-dependent and velocity-dependent contribution, the so-called Yafet shift of the position~\eqref{eq:Yafet}. This leads to another anomalous contribution to the velocity operator~\eqref{eq:velocity_so_2} and to the matrix current~\eqref{eq:currentmatrix}.
    
    Here, we will discuss these anomalous current contributions to the side-jump mechanism and compare it to the contribution from the side-jump self-energy obtained previously.
    
 \subsection{Anomalous Contributions to the Matrix Current}
  \label{subsec:sidejump_anomalous}
The shift in the position operator (\ref{eq:position_operator}) leads to a shift in the velocity operator,    
    \begin{equation*}
      \hat{\vect{v}}
      =
      \dot{\hat{\vect{r}}}_\text{eff}
      =
      \dot{\vect{r}}
      +
      \dot{\hat{\vect{r}}}_\text{so}.
    \end{equation*}
		The velocity operator is calculated from the Heisenberg equation of motion in terms of the Hamiltonian of Eq.~\eqref{eq:hamiltonian} and acquires two spin-dependent corrections as a consequence of spin-orbit coupling. The first emerges from
    \begin{equation*}
      \dot{\vect{r}}
      =
      -
      \im
      \big[
	\vect{r}, \hat{\uppercase{\mathcal{H}}}
      \big]_{-}
      =
      \vect{v} - \frac{e}{m}\hat{\tau}_3\vec{A}(1)
      +
      \hat{\vect{v}}_\text{so}^{(1)}
      ,
    \end{equation*}
    where
    \begin{subequations}
    \label{eq:velocity_so}
    \begin{equation}
    \label{eq:velocity_so_1}
      \hat{\vect{v}}_\text{so}^{(1)}
      =
      -
      \im
      \big[
	\vect{r}, \hat{U}_\text{so}
      \big]_{-}
      =
      \gamma      
      \sum_i
      \big(
       \tauhat{3} \hat{\vect{\alpha}}
       \times
       \vect{\nabla} u(\vect{r} - \vect{r}_i)
      \big)
      .
    \end{equation}
    The second correction $\hat{\vect{v}}_\text{so}^{(2)} = \dot{\hat{\vect{r}}}_\text{so}$ arises from the Yafet shift of the position operator~\eqref{eq:Yafet} and is, to the first order in the spin-orbit coupling strength,
    \begin{equation}
    \label{eq:velocity_so_2}
      \hat{\vect{v}}_\text{so}^{(2)}
      =
      -
      \im
      \big[
	\hat{\vect{r}}_\text{so}, \hat{U}
      \big]_{-}
      =
      \gamma      
      \sum_i
      \big(
       \tauhat{3} \hat{\vect{\alpha}}
       \times
       \vect{\nabla} u(\vect{r} - \vect{r}_i)
      \big)
      .
    \end{equation}
    \end{subequations}
    Note that $\hat{\vect{v}}_\text{so}^{(1)}$ and $\hat{\vect{v}}_\text{so}^{(2)}$ are identical, giving rise to an overall factor of $2$. In total, the velocity operator is thus given by
      \begin{equation}
      \label{eq:velocity_definition}
	\hat{\vect{v}}(\vect{r})
	=
	- \frac{\im}{m} \vect{\partial}_{\vect{r}}
	+
\hat{\vect{v}}_\text{so}(\vect{r})
	.
      \end{equation}
    Note that the spin current density in this definition is not conserved in the presence of magnetic impurities or spin-orbit coupling.
    As discussed in Sec.~\ref{sec:derivation}, the velocity operator~\eqref{eq:velocity_definition} acquires the two spin-dependent corrections of Eq.~\eqref{eq:velocity_so} as a consequence of the spin-orbit coupling, giving rise to the overall anomalous velocity contribution
    \begin{equation}
    \label{eq:sidejump_anomalous_total}
      \hat{\vect{v}}_\text{so}(\vect{r})
      =
      2 \gamma
      \sum_i
      \big(
	\tauhat{3} \hat{\vect{\alpha}} \times \vect{\nabla} u(\vect{r} - \vect{r}_i)
      \big)
      .
    \end{equation}
    This anomalous contribution to the matrix current defined in Eq.~\eqref{eq:currentmatrix} reads as
    \begin{equation}
    \label{eq:sidejump_anomalous_current}
    \begin{split}
      & \;
      \hat{\vect{\jmath}}_\text{so}(\cone{1})
      \\ 
      =
      & \;
      \frac{\im}{2 N_0}
      \lim_{\cone{1'} \rightarrow \cone{1}}
      \big(
	\hat{\vect{v}}_\text{so}(\vect{r}_\cone{1}) \hat{G}^\text{K}(\cone{1}, \cone{1'})
	+
	\hat{G}^\text{K}(\cone{1}, \cone{1'}) \hat{\vect{v}}_\text{so}(\vect{r}_\cone{1'})
      \big)
      .
    \end{split}
    \end{equation}
    The challenge in evaluating this expression is computing the impurity average. While the conventional velocity operator is independent of the impurity configuration, the anomalous contribution explicitly depends on the impurities and we need to evaluate $\langle \hat{\vect{v}}_\text{so} \hat{G}^\text{K} \rangle_\text{c}$ and $\langle \hat{G}^\text{K} \hat{\vect{v}}_\text{so} \rangle_\text{c}$. This can be achieved by following the procedure in Ref.~\onlinecite{Shchelushkin2005}: from the Dyson equation, it follows that $\langle \hat{U}_\text{tot} \check{G} \rangle_\text{c} = \check{\Sigma} \check{G}_\text{c}$ and $\langle \check{G} \hat{U}_\text{tot} \rangle_\text{c} = \check{G}_\text{c} \check{\Sigma}$, where $\check{\Sigma}$ is the self-energy. Consequently
    \begin{subequations}
    \label{eq:sidejump_selfenergy_anom_definition}
    \begin{align}
    \label{eq:sidejump_selfenergy_anom_definition_left_imp}
      \big\langle
	\hat{\vect{v}}_\text{so}(\vect{r}_\cone{1})
	\hat{G}^\text{K}(\cone{1}, \cone{1'})
      \big\rangle_\text{c}
      &
      =
      \int \diffint{\cone{2}}{}
      \Big(
	\check{\vect{\Sigma}}_\text{sj}^\text{(l)}(\cone{1}, \cone{2})
	\check{G}_\text{c}(\cone{2}, \cone{1'})
      \Big)^\text{K}
      ,
      \\
    \label{eq:sidejump_selfenergy_anom_definition_right_imp}
      \big\langle
	\hat{G}^\text{K}(\cone{1}, \cone{1'})
	\hat{\vect{v}}_\text{so}(\vect{r}_\cone{1'})
      \big\rangle_\text{c}
      &
      =
      \int \diffint{\cone{2}}{}
      \Big(
	\check{G}_\text{c}(\cone{1}, \cone{2})
	\check{\vect{\Sigma}}_\text{sj}^\text{(r)}(\cone{2}, \cone{1'})
      \Big)^\text{K}
      ,
    \end{align}
    where, within the self-consistent Born approximation to the first order in the spin-orbit coupling, see Fig.~\ref{fig:selfenergy_born_1},
    \begin{align}
    \notag
      & \;
      \check{\vect{\Sigma}}_\text{sj}^\text{(l)}(\cone{1}, \cone{1'})
      \\
    \label{eq:sidejump_selfenergy_anom_definition_left}
      =
      & \;
      2 \gamma n
      \int \diffint{\vect{r}_i}{}
      \big(
	\tauhat{3} \hat{\vect{\alpha}} \times \vect{\nabla} u(\vect{r}_\cone{1} - \vect{r}_i)
      \big)
      \check{G}_\text{c}(\cone{1}, \cone{1'})
      u(\vect{r}_\cone{1'} - \vect{r}_i)
      ,
      \\ 
    \notag
      & \;
      \check{\vect{\Sigma}}_\text{sj}^\text{(r)}(\cone{1}, \cone{1'})
      \\
    \label{eq:sidejump_selfenergy_anom_definition_right}
      =
      & \;
      2 \gamma n
      \int \diffint{\vect{r}_i}{}
      u(\vect{r}_\cone{1} - \vect{r}_i)
      \check{G}_\text{c}(\cone{1}, \cone{1'})
      \big(
	\tauhat{3} \hat{\vect{\alpha}} \times \vect{\nabla} u(\vect{r}_\cone{1'} - \vect{r}_i)
      \big)
      .
    \end{align}
    \end{subequations}
    In the mixed representation, we have
    \begin{subequations}
    \label{eq:sidejump_selfenergy_anom}
    \begin{align}
    \notag
      & \;
      \check{\vect{\Sigma}}_\text{sj}^\text{(l)}(\vect{R}, \vect{p}, \epsilon)
      \\
    \label{eq:sidejump_selfenergy_anom_left}
      =& \;
      2 \im \gamma n
      \int \! \frac{\mathrm{d}\vect{q}}{(2 \pi)^3}\,
      |u(\vect{p} - \vect{q})|^2
      \tauhat{3} \hat{\vect{\alpha}} \times (\vect{p} - \vect{q}) \check{G}_\text{c}(\vect{R}, \vect{q}, \epsilon)
      ,
      \\ 
    \notag
      & \;
      \check{\vect{\Sigma}}_\text{sj}^\text{(r)}(\vect{R}, \vect{p}, \epsilon)
      \\
    \label{eq:sidejump_selfenergy_anom_right}
      =& \;
      - 2 \im \gamma n
      \int \! \frac{\mathrm{d}\vect{q}}{(2 \pi)^3}\,
      |u(\vect{p} - \vect{q})|^2
      \check{G}_\text{c}(\vect{R}, \vect{q}, \epsilon) \tauhat{3} \hat{\vect{\alpha}} \times (\vect{p} - \vect{q})
      ,
    \end{align}
    \end{subequations}
    and in the quasiclassical approximation, we obtain
    \begin{subequations}
    \label{eq:sidejump_selfenergy_anom_quasi}
    \begin{align}
    \label{eq:sidejump_selfenergy_anom_quasi_left}
      \check{\vect{\sigma}}_\text{sj}^\text{(l)}(\vect{p}_\text{F})
      &=
      \frac{\gamma p_\text{F}}{\tau_\text{tr}}
      \big(
	\tauhat{3} \hat{\vect{\alpha}} \times \vect{e}_p
      \big)
      \big(
	\check{g}_\text{s}
	-
	(\vect{e}_p \cdot \check{\vect{g}})
      \big)
      ,
      \\
    \label{eq:sidejump_selfenergy_anom_quasi_right}
      \check{\vect{\sigma}}_\text{sj}^\text{(r)}(\vect{p}_\text{F})
      &=
      -
      \frac{\gamma p_\text{F}}{\tau_\text{tr}}
      \big(
	\check{g}_\text{s}
	-
	(\vect{e}_p \cdot \check{\vect{g}})
      \big)
      \big(
	\tauhat{3} \hat{\vect{\alpha}} \times \vect{e}_p
      \big)
      ,
    \end{align}
    \end{subequations}
    where we used the expansion $\check{g}(\vect{R}, \vect{q}_\text{F}, \epsilon) \approx \check{g}_\text{s}(\vect{R}, \epsilon) + \vect{e}_q \cdot \vect{\check{g}}(\vect{R}, \epsilon)$ and performed the angular average over $\vect{q}$. We also omitted $\vect{R}$ and $\epsilon$ for brevity.
    
    Using Eq.~\eqref{eq:sidejump_selfenergy_anom_definition}, the anomalous contribution to the impurity-averaged matrix current~\eqref{eq:sidejump_anomalous_current} in the Fourier representation is
    \begin{widetext}
    \begin{equation}
    \label{eq:sidejump_anomalous_current_averaged}
      \big\langle\hat{\vect{\jmath}}_\text{so}\big\rangle_\text{c}(\vect{R})
      =
      \frac{\im}{4 \pi N_0}
      \int \diffint{\epsilon}{}
      \int \! \frac{\mathrm{d}\vect{p}}{(2 \pi)^3}\,
      \Big(
	\check{\vect{\Sigma}}_\text{sj}^\text{(l)}(\vect{R}, \vect{p}, \epsilon)
	\check{G}_\text{c}(\vect{R}, \vect{p}, \epsilon)
	+
	\check{G}_\text{c}(\vect{R}, \vect{p}, \epsilon)
	\check{\vect{\Sigma}}_\text{sj}^\text{(r)}(\vect{R}, \vect{p}, \epsilon)
      \Big)^\text{K}
      ,
    \end{equation}
    to the lowest order in the gradient approximation. In the quasiclassical approximation, this becomes
		
    \begin{equation}
    \label{eq:sidejump_anomalous_current_quasi}
      \big\langle\hat{\vect{\jmath}}_\text{so}\big\rangle_\text{c}(\vect{R})
      =
      \frac{1}{4}
      \int \diffint{\epsilon}{}
      \Big\langle
      \big(
	\check{\vect{\sigma}}_\text{sj}^\text{(l)}(\vect{R}, \vect{p}, \epsilon)
	\check{g}(\vect{R}, \vect{p}, \epsilon)
	+
	\check{g}(\vect{R}, \vect{p}, \epsilon)
	\check{\vect{\sigma}}_\text{sj}^\text{(r)}(\vect{R}, \vect{p}, \epsilon)
      \big)^\text{K}
      \Big\rangle_\text{F}
      ,
    \end{equation}
    \end{widetext}
    where $\langle \dots \rangle_\text{F} = \int \! \frac{\mathrm{d}\vect{e}_p}{4 \pi}\, \dots$ denotes an angular average over all momentum directions at the Fermi surface. We can now use the expansion of the Green's functions in spherical harmonics, Eqs.~\eqref{eq:greens_harmonics} and~\eqref{eq:normalization_harmonics}; insert the results of Eq.~\eqref{eq:sidejump_selfenergy_anom_quasi}; and perform the angular average. Note that, in general, an additional term emerges when computing Eq.~\eqref{eq:sidejump_selfenergy_anom_quasi}. However, this term vanishes when the angular average is performed on Eq.~\eqref{eq:sidejump_anomalous_current_quasi} and is consequently of no interest. With this, we finally obtain the anomalous correction to the matrix current in Eq.~\eqref{eq:currentmatrix}.

  \subsection{Contributions to the Spin Hall Effect}  
  \label{subsec:sidejump_total}
  
    As mentioned, the origins of the side-jump mechanism are three-fold\footnotemark[5]. An additional self-energy contribution~\eqref{eq:selfenergy_sj} emerges when evaluating the self-energy to the first order in the spin-orbit interaction strength~\eqref{eq:selfenergy_definition_so1} to next-to-leading order in the gradient approximation. This self-energy contributes to the spin Hall effect in Eq.~\eqref{eq:currents_sh} with a term proportional to $\gamma m / \tau_\text{tr}$. Additionally, the current acquires the two spin-dependent corrections of Eq.~\eqref{eq:velocity_so}. These anomalous corrections contribute with a term proportional to $\gamma m / \tau_\text{tr}$ each. In total, the side-jump contribution to the spin Hall effect is thus given by $\chi_\text{sH}^\text{(sj)} = 3 \gamma m / \tau_\text{tr}$.

\nocite{*}


\begin{thebibliography}{58}
\bibitem{comment}
Below, we consider contributions to the side-jump mechanism to the spin Hall effect arising from the next-to-leading order in the gradient approximation. While similar terms can also arise from next-to-leading order in this expression, we demonstrate below that all contributions (side-jump, skew, and Yafet) to the spin Hall angle are renormalized in the same way in the superconducting state. Hence, based on this symmetry, we expect that possible additional contributions can renormalize the side-jump coefficient but not how it is affected by superconducing correlations.
\end{thebibliography}
\end{document}